\documentclass[fleqn,usenatbib]{mnras}
\usepackage{amsmath}
\usepackage{amssymb}
\usepackage{graphicx}
\usepackage{float}
\usepackage[T1]{fontenc}
\usepackage{multirow}

\usepackage{xcolor}
\colorlet{purple}{blue!50!red}
\colorlet{orange}{red!40!yellow}

\begin{document}

\title[Inferring subhalo effective density slopes from strong lensing observations with machine learning]{Inferring subhalo effective density slopes from strong lensing observations with neural likelihood-ratio estimation}
\author[G. Zhang, S. Mishra-Sharma, and C. Dvorkin]{Gemma Zhang$^{1}$\thanks{\href{mailto:yzhang7@g.harvard.edu}{yzhang7@g.harvard.edu}}, Siddharth Mishra-Sharma$^{1,2,3}$, and Cora Dvorkin$^{1}$
\\\\
$^{1}$Department of Physics, Harvard University, Cambridge, MA 02138, USA\\
$^{2}$The NSF AI Institute for Artificial Intelligence and Fundamental Interactions\\
$^{3}$Center for Theoretical Physics, Massachusetts Institute of Technology, Cambridge, MA 02139, USA\\
}

\label{firstpage}
\pagerange{\pageref{firstpage}--\pageref{lastpage}}
\maketitle

\begin{abstract}
    Strong gravitational lensing has emerged as a promising approach for probing dark matter models on sub-galactic scales. Recent work has proposed the subhalo effective density slope as a more reliable observable than the commonly used subhalo mass function. The subhalo effective density slope is a measurement independent of assumptions about the underlying density profile and can be inferred for individual subhalos through traditional sampling methods. To go beyond individual subhalo measurements, we leverage recent advances in machine learning and introduce a neural likelihood-ratio estimator to infer an effective density slope for populations of subhalos. We demonstrate that our method is capable of harnessing the statistical power of multiple subhalos (within and across multiple images) to distinguish between characteristics of different subhalo populations. The computational efficiency warranted by the neural likelihood-ratio estimator over traditional sampling enables statistical studies of dark matter perturbers and is particularly useful as we expect an influx of strong lensing systems from upcoming surveys. 
\end{abstract}


\section{Introduction}

Dark matter~(DM) accounts for approximately 84\% of the total matter budget of the Universe, yet its nature remains one of the biggest unanswered questions in fundamental physics~\citep{planck2020planck}. The current standard $\Lambda$CDM cosmological model stipulates that dark matter is cold and collisionless and predicts a bottom-up hierarchical formation of structures. 
The $\Lambda$CDM model has been extensively tested by measurements at and above galactic scales, but evidence remains scarce on sub-galactic scales where many DM models differ in their predictions~\citep{bode2001wdm, anderhalden2013hints, bullock2017smallscale, buckley2018gravitational}. Much of the challenge in testing the small-scale structure of dark matter lies in complications stemming from accurately modeling baryonic effects in galaxies. Low-mass subhalos ($\lesssim 10^9$ $\mathrm M_{\odot}$) residing within a larger host dark matter halo provide a promising laboratory to test DM models because they are typically devoid of luminous matter and are thus less affected by baryonic processes~\citep{fitts2017fire, read2017stellar, kim2018missing}. 

Because of the lack of luminous matter in these subhalos, observations of their gravitational effects constitute one of the only ways of characterizing them. Within the Local Group, subhalos can be probed by analyzing stellar streams~\citep{carlberg2016modeling, bonaca2020highres, banik2021novel}, stellar wakes~\citep{buschmann2018stellar}, and astrometric measurements~\citep{feldmann2015detecting, vantilburg2018halometry, mishrasharma2020powerhalometry, mondino2020first}. Beyond the immediate neighborhood of the Milky Way, strong gravitational lensing has been the most common method for constraining subhalo properties. Strong gravitational lensing occurs when a distant source, which can be an extended galaxy or a point-like quasar, is in close alignment with a foreground massive structure. In systems with quasar sources, flux ratio measurements are typically used to detect subhalos and constrain their properties~\citep{mao1998evidence, metcalf2001compound, moustakas2003detecting, fadely2012substructure, macleod2013detection, nierenberg2014detection, nierenberg2017probing, gilman2020wdm, gilman2021strong}. In this work, we will focus on strong gravitational lensing systems where both the source and lens are galaxies. So far, $\mathcal{O}$(100) such systems have been observed across existing datasets~\citep{metcalf2019strong, brownstein201bells, bolton2008slacs, jacobs2019des, storfer2022desi}, and several systems contain claimed substructure detections~\citep{vegetti2010detection, vegetti2012detection, hezaveh2016detection}. With forthcoming data from large-scale imaging surveys, the number of detected strong lensing systems is expected to increase by orders of magnitude~\citep{laureijs2011euclid,Collett:2015roa,mckean2015ska, bechtol2019lsst, jacobs2019des, huang2021desi, storfer2022desi}, so it is timely to think about the prospects for using strong gravitational lensing for population-level studies of dark matter. 

Previous analyses of substructure fall into two categories: direct detection in which individual subhalos are constrained through lens modeling~\citep{vegetti2014inferencecdmsubmf, ritondale2019lowmass}, and statistical detection, which attempts to extract the collective effect of subhalos~\citep{dalal2001direction, hezaveh2014measuring, cyr-Racine2015darkcensus, diazRivero2017powerspec, diazRivero2018powerspecethos, birrer2017lensingsubs, daylan2018probing, brennan2018quantifying, gilman2018, cyrracine2019, gilman2020wdm, he2022}. On the statistical detection front, machine learning has been an emerging approach for extracting population-level statistics of substructure~\citep{brewer2016transdim, brehmer2019mining, Ostdiek:2020mvo, Ostdiek:2020cqz,carenawagner2022images}. However, these works typically rely on constraining the subhalo mass function (SHMF), which depends on assumptions about the subhalo density profile in areas that are not directly probed. This can hinder our ability to reliably constrain the subhalo mass function. As a result, \citet{2017ApJ...845..118M} proposed a robust subhalo mass observable, called an effective subhalo lensing mass, and showed that this mass definition is independent of assumptions about the subhalo density profile. Moreover, relying on subhalo mass measurements overlooks information hidden in the subhalo central density profiles, which have been shown to be an important complementary probe of DM to the subhalo mass~\citep{minor2021inferring, minor2021unexpected, amorisco2022}. To address this, \citet{sengul2022probing} proposed a new observable, which they call the ``effective density slope,'' as a robust way of distinguishing between different classes of DM models. This effective density slope is defined as the slope of the density profile of a subhalo at scales where it has the largest observable effect. \citet{sengul2022probing} showed that the effective slope of an individual subhalo can be accurately constrained using traditional sampling techniques. While characterizing individual subhalos is useful, it is often limited to leveraging the effect of more massive subhalos in a system. With the inclusion of more subhalos, the joint parameter space of individual subhalo properties increases significantly and sampling becomes prohibitive. This makes it particularly challenging to obtain enough measurements to make statistically robust inference. Moreover, characterizing individual subhalos does not account for the (expected) effect of a large population of unresolved lower-mass subhalos.

Modern machine learning developments have made it possible to overcome these challenges. Simulation-based inference using machine learning tools has been applied to an array of problems in cosmology~\citep{PerreaultLevasseur2017uncertainties, delfi1,delfi2,Alsing2019fastlldfree, Coogan2020targeted, Legin:2021zup, Wagner-Carena2020hierarchical, Makinen:2021nly, Dai2022trenf,Cole:2021gwr}. In this paper, we propose a machine learning approach that makes use of neural likelihood-ratio estimation~\citep{cranmer2015approximating, baldi2016parameterized} in order to probe a population-level effective density slope. 
A crucial property of the method is that it can be used to efficiently implicitly marginalize over the large latent space in the lensing model without explicitly inferring parameters associated with individual subhalos. The advantage of this approach is two-fold: firstly, the number of subhalos is expected to increase as the mass decreases, so probing a population-level statistic takes into account the collective effect of low-mass subhalos; secondly, after investing in an initial training overhead, the machine learning model is computationally efficient at combining information across a large sample of lenses in a dataset. 

This paper is organized as follows. In Sec.~\ref{sec:data}, we lay out the details of the lensing forward model used to generate the mock lensed images used for training as well as validation. In Sec.~\ref{sec:model}, we describe the neural likelihood-ratio estimation technique as well as model architecture used in this work. In Sec.~\ref{sec:results}, we demonstrate our model's ability to distinguish between subhalo populations with different density slopes and compare the model's predicted slopes with the analytically derived effective density slopes following \citet{sengul2022probing}. Finally, in Sec.~\ref{sec:conlcusion}, we conclude with a summary of our results and discuss potential improvements. 

\section{Data generation}
\label{sec:data}

To generate the strong lensing images for model training and testing, we use the software package \textsc{lenstronomy}~\citep{birrer2015gravitational, birrer2018lenstronomy}. We generate ($100 \times 100$)~pixel$^2$ images with a resolution of $0.08''$ per pixel side. This corresponds to a field of view of $8'' \times 8''$ in an image. Generating an image requires several ingredients: a source galaxy, a main (host) lens galaxy, a population of subhalos, and a specification of the instrumental configuration. We discuss each of these components in detail below and the concrete values of parameters used for data generation are summarized in Table~\ref{table:params}. 

\subsection{Source and main lens}

The source and the main lens are the primary ingredients producing galaxy-galaxy lensing images. In each lensing system, light rays emitted by the source galaxy get gravitationally deflected by the main lens en route to the detector. In strong gravitational lensing, the bending of source light often results in characteristic lensed arcs in the observed images. In our datasets, we place the source at redshift $z_{\mathrm{source}} = 1.0$ and lens at redshift $z_{\mathrm{lens}} = 0.5$. 

To simulate the source light in our images, we use source galaxy images provided by the \textsc{paltas} pipeline~\citep{carenawagner2022images}. These sources are real galaxy images taken by the Hubble Space Telescope~(HST) Cosmic Evolution Survey~(COSMOS)~\citep{scoville2007cosmos, koekemoer2007cosmos}. The \textsc{paltas} package takes a sub-sample of the HST COSMOS survey galaxies from the GREAT3 gravitational lensing challenge dataset~\citep{mandelbaum2012precision, mandelbaum2014great3}. It then filters out 2,262 source candidates by putting constraints on parameters such as the pixel resolution and galaxy apparent magnitude and manually inspecting to keep only well-resolved images. Out of the 2,262 source galaxies available, we reserve 2,163 for modeling the training images, 70 for validation, and the remainder for testing and evaluation. This ensures that we can test the ability of our model to generalize to sources that it has not seen during training. To model the source light for a lensing image, we randomly draw a galaxy from the COSMOS catalog, apply a random rotation, and then draw the source position coordinates $x_{\mathrm{source}}, y_{\mathrm{source}}$ from a uniform interval of $[-0.1, 0.1]''$ with the center of the image being at $[0, 0]$. 

We model the main lens using a singular isothermal ellipsoid~(SIE) profile~\citep{kormann1994isothermal}, whose convergence is parameterized in \textsc{lenstronomy} as 
\begin{equation}
    \kappa(x, y) = \frac{1}{2}\left(\frac{\theta_E}{\sqrt{q x_{\phi}^2 + y_{\phi}^2/q}}\right),
\end{equation}
where $\theta_E$ is the Einstein radius, $q$ is the minor/major axis ratio, and $x_{\phi}, y_{\phi}$ are positions on the axes aligning with the major and minor axes of the lens. The angle of rotation $\phi$ is the angle between the major and minor axes and the fixed $x, y$ axes of the image. The inputs into \textsc{lenstronomy} are the ellipticity moduli instead of the major/minor axis ratio, and they are related as follows: 
\begin{align}
    e_1 &= \frac{1 - q}{1 + q} \cos(2\phi), \\ 
    e_2 &= \frac{1 - q}{1 + q} \sin(2\phi).
\end{align}
We also include an external shear parameterized by $\gamma_{\mathrm{shear}, 1}$ and $\gamma_{\mathrm{shear}, 2}$ as part of our lens model~\citep{Keeton1996shear}. The shear parameters $\gamma_{\mathrm{shear}, 1}$ and $\gamma_{\mathrm{shear}, 2}$ are the diagonal and off-diagonal terms of the shear matrix, respectively. The external shear encodes the stretching of the source shape by nearby or line-of-sight massive structures. In modeling each image, we vary the center position, the Einstein radius, the shear parameters, and the ellipticity of the main lens as specified in Table~\ref{table:params}. Our choice of the Einstein radius corresponds to an average main lens mass of approximately $1.6 \times 10^{13} \mathrm M_{\odot}$. 

\begin{table*}
    \centering
    \begin{tabular}{l c} 
        \hline
        \bf{Parameter} & \bf{Distribution} \\
        \hline 
        \multicolumn{2}{l}{\underline{Source}} \\
        Source redshift & $z_{\mathrm{source}} = 1.0$ \\
        $x$-coordinate & $x_{\mathrm{source}} \sim \mathcal U(-0.1'', 0.1'')$ \\
        $y$-coordinate & $y_{\mathrm{source}} \sim \mathcal U(-0.1'', 0.1'')$ \\
        \hline 
        \multicolumn{2}{l}{\underline{Main lens}} \\
        Lens redshift & $z_{\mathrm{lens}} = 0.5$ \\
        $x$-coordinate & $x_{\mathrm{lens}} \sim \mathcal U(-0.2'', 0.2'')$ \\
        $y$-coordinate & $y_{\mathrm{lens}} \sim \mathcal U(-0.2'', 0.2'')$ \\
        Einstein radius & $\theta_E \sim \mathcal U(2.7'', 3'')$ \\
        \multirow{2}{*}{Ellipticities} & $e_1 \sim \mathcal U(-0.2, 0.2)$ \\
        & $e_2 \sim \mathcal U(-0.2, 0.2)$ \\
        \multirow{2}{*}{External shear} & $\gamma_{\mathrm{shear}, 1} \sim \mathcal U(-0.1, 0.1)$ \\
        & $\gamma_{\mathrm{shear}, 2} \sim \mathcal U(-0.1, 0.1)$ \\
        \hline 
        \multicolumn{2}{l}{\underline{Subhalos}} \\
        \multirow{2}{*}{EPL ellipticities} & $e_1 \sim \mathcal U(-0.2, 0.2)$ \\
        & $e_2 \sim \mathcal U(-0.2, 0.2)$ \\
        EPL slope of density profile per lens system & $\gamma \sim \mathcal U(1.1, 2.9)$ \\
        EPL slope of density profile per subhalo & $\gamma_i \sim \mathcal{N}(\gamma, 0.1\gamma)$ \\
        Subhalo mass function power-law slope & $-1.9$ \\
        Subhalo mass & $M_{200} \in [10^7, 10^{10}]\mathrm M_{\odot}$; $[10^8, 10^{10}]\mathrm M_{\odot}$; $10^9 \mathrm M_{\odot}$ \\
        Number of subhalos per image & $N_{\mathrm{sub}} \sim \mathcal U\{0, 500\}$; $\mathcal U\{0, 60\}$; $\mathcal U\{0, 100\}$ \\
        \hline 
    \end{tabular}
    \caption{Parameters of the main components of a strong gravitational lensing system and their respective training distribution in our mock images. The test set follows the same parameter distributions except for when the subhalos have NFW instead of EPL profiles, as discussed in Sec.~\ref{subsubsec:test_subhalo}.}
    \label{table:params}
\end{table*}

\subsection{Subhalos}

We refer to a lensed image resulting from a source and a main lens deflector as a ``smooth model image.'' The presence of subhalos near the bright lensed arcs of a smooth model image can leave observable signatures in the lensed image. Thus, by learning to detect the effect of these perturbations in the lensed images, we can probe the properties of subhalos. 

Only subhalos in the bright regions of a smooth model image have potentially observable effects because the lensing effects of subhalos elsewhere are degenerate with variations in the main lens mass. As such, in our datasets, we select pixels whose brightness is more than half of the maximum brightness in the smooth model image and then randomly place subhalos only in these areas. For our main results in Sec.~\ref{sec:results}, we draw subhalos with virial mass $M_{200}$ between $10^7~\mathrm{M}_{\odot}$ and $10^{10}~\mathrm{M}_{\odot}$ from a power-law subhalo mass function given by $dN_{\mathrm{sub}}/dM_{200} \propto M_{200}^{-1.9}$ where $N_{\mathrm{sub}}$ is the number of subhalos and $M_{200}$ is the total mass of the subhalo within $r_{200}$, which is the radius within which the average mass density of the subhalo is 200 times the critical density of the Universe.

The number of subhalos we add to each image is uniformly distributed between 0 and 500. This range was chosen based on existing estimates of $f_{\mathrm{sub}}$, which denotes the fraction of substructure mass relative to the host galaxy mass. Concrete constraints on $f_{\mathrm{sub}}$ in existing literature are relatively scarce and can vary significantly between galaxies~\citep{CaganSengul:2020nat}. Using the SLACS galaxies, \citet{vegetti2014inferencecdmsubmf} gives an upper bound on $f_{\mathrm{sub}}$ of approximately 0.007 for subhalos between $4\times 10^6\mathrm M_{\odot}$ and $4\times 10^9\mathrm M_{\odot}$ in a 32~kpc$^2$ region around the Einstein ring at $z_{\mathrm{lens}} = 0.2$; this translates to an approximate upper bound of 800 subhalos between $10^7\mathrm M_{\odot}$ and $10^{10}\mathrm M_{\odot}$ in the bright regions in our images at $z_{\mathrm{lens}} = 0.5$ for a main lens of $1.6 \times 10^{13}\mathrm M_{\odot}$. On the other hand, \citet{ritondale2019lowmass} uses the BELLS GALLERY galaxies and gives an upper bound on $f_{\mathrm{sub}}$ of roughly 0.07 for subhalos between $10^5\mathrm M_{\odot}$ and $10^{11}\mathrm M_{\odot}$ in the whole image; this translates to an approximate upper bound of 250 subhalos between $10^7\mathrm M_{\odot}$ and $10^{10}\mathrm M_{\odot}$ in the bright regions in our images at $z_{\mathrm{lens}} = 0.5$ for a main lens of $1.6 \times 10^{13}\mathrm M_{\odot}$. For our datasets, we choose a rough average between these two estimates: an upper bound on $N_{\mathrm{sub}}$ of 500. In Sec.~\ref{sec:results}, for demonstrative and comparison purposes, we also show additional results for a few different mass ranges and numbers of subhalos, and we specify these ranges when applicable. 

All of our mock images share the same aforementioned parameters. However, for practical purposes, we model the subhalo density profiles in the training/validation set and the test set differently and discuss them in detail below. 

\subsubsection{Training and validation sets}

To produce the training and validation sets, we add subhalos that follow the elliptical power law~(EPL) profile~\citep{barkana1998epl}, whose convergence at position $(x, y)$ on the lens plane is defined as follows: 
\begin{align}
    \kappa(x, y) = \frac{3 - \gamma}{2} \left(\frac{\theta_E}{\sqrt{qx_{\phi}^2 + y_{\phi}^2/q}}\right)^{\gamma - 1},
    \label{eq:epl_kappa}
\end{align}
where $\theta_E, q, x_{\phi}, y_{\phi}$ are defined similarly to those in the SIE profile, and $\gamma$ is an additional free parameter that controls the slope of the density profile of a subhalo. For $q = 1$, the average convergence $\bar{\kappa}$ within a given radius $r$ from the center of an EPL subhalo is
\begin{align}
     \bar{\kappa} = \frac{2\pi\int_0^r r' \kappa(r') dr'}{\pi r^2} \propto r^{1 -\gamma}, 
\end{align}
so that its power-law slope is given as a function of the average convergence as 
\begin{align}
    \gamma = 1 - \frac{d \ln\bar{\kappa}}{d \ln r}.
    \label{eq:slope_kappabar}
\end{align}
This particular profile choice allows us to conveniently vary the slope of the density profile and label each training image with its true underlying slope. Although EPL is not a sufficiently realistic profile for subhalos, this choice is deliberate because our model requires the true parameter value as an input during training, as will be discussed in Sec.~\ref{subsec:train_details}. 

The 3D density of EPL subhalos scales with the radial distance as $\propto r^{-\gamma}$. Using Eq.~\eqref{eq:epl_kappa} to solve for the normalization of the 3D density, we see that, assuming $q = 1$ for no ellipticity, the radial density profile of an EPL subhalo is 
\begin{align}
    \rho(r) = \rho_{\mathrm{EPL}} r^{-\gamma}, 
\end{align}
where $\rho_{\mathrm{EPL}} = \frac{2\sqrt{\pi}}{3 - \gamma} \frac{\Gamma(\frac{\gamma - 1}{2})}{\Gamma(\frac{\gamma}{2})}$ is a constant with $\Gamma$ being the Gamma function. Note that when $\gamma = 1$, the mass increases radially and diverges eventually, and when $\gamma = 3$, the density normalization $\rho_{\mathrm{EPL}}$ diverges. As a result, we constrain $\gamma$ to values between 1 and 3 in order to ensure that the subhalo masses are physical. To model the subhalos in each image, we first draw a slope from the uniform distribution  $\gamma \sim \mathcal U(1.1, 2.9)$ and then draw normally distributed slopes $\gamma_i \sim \mathcal{N}(\gamma, 0.1\gamma)$ for the $i$th subhalo. We truncate the $\mathcal{N}(\gamma, 0.1\gamma)$ distribution so that $\gamma_i$ is constrained to $(1, 3)$. Drawing the slope of each subhalo from a normal distribution more realistically simulates the natural variations in slope measurements that are observed in simulations~\citep{sengul2022probing}. 

\subsubsection{Test set}
\label{subsubsec:test_subhalo}

In the test set, we model subhalos with profiles that are more realistic than an EPL profile. Here we use the Navarro–Frenk–White (NFW) profile~\citep{navarro1997nfw}, whose radial density profile is defined as 
\begin{align}
    \rho(r) = \frac{\rho_0}{\frac{r}{r_s}\left(1 + \frac{r}{r_s}\right)^2},
\end{align}
where $r$ is the radial distance from the center of the subhalo, and the normalization $\rho_0$ and the scale radius $r_s$ are parameters that vary between subhalos. An NFW profile can alternatively be parameterized by $M_{200}$ and concentration $c_{200}$. The concentration is defined as the ratio between $r_{200}$ and the scale radius: $r_{200} = c_{200} r_s$. To produce our main test set, we draw masses from the subhalo mass function and set its $c_{200}$ using a fixed mass-concentration relation extrapolated from \citet{dutton2014cdmhalo}, which we denote as $f_{\mathrm{CDM}}(M_{200})$. To simulate subhalo populations with different slopes, we vary the subhalo concentrations in the test set using an overall multiplicative factor $a$, such that the modified mass-concentration becomes $a$$\times$$f_{\mathrm{CDM}}(M_{200})$. Subhalo populations with $a < 1$ have concentrations lower than those given by $f_{\mathrm{CDM}}(M_{200})$ and have more cored inner density profiles while subhalo populations with $a > 1$ have more cuspy inner density profiles. 

Adding subhalos in the lens plane increases the total mass that lenses the source light, thereby enlarging the Einstein radii in the images produced. In some of our experiments in Sec.~\ref{subsec:nfw_concentration}, we add a large total subhalo mass in each lensing system, which risks producing an effective Einstein radius larger than the image size. To mitigate this, we add a sheet of constant negative convergence to our modeling so that the Einstein radius of an image including subhalos gets restored to its value in the smooth model image. To add the sheet of negative convergence to an image, we compute the total convergence of all the subhalos and add a constant negative convergence across all pixels in the image so that they cancel out the total subhalo convergence. 

\subsection{Instrumental details}

To create the lensed images, we use the WFC3 F160W camera band configuration and a Gaussian point spread function~(PSF) with a full width at half maximum of $0.08''$. We use the \textsc{lenstronomy} HST configurations with a sky brightness of 22.3 and magnitude zero point of 25.96. In each image, we add 10 orbits of HST noise, with an exposure time of 5,400~seconds per orbit. 

\section{Model and inference}
\label{sec:model}

For our task, we train a neural likelihood-ratio estimator using parameterized classifiers~\citep{cranmer2015approximating, baldi2016parameterized,hermans2019likelihoodfree}, a technique that has been previously proposed in the context of astrophysical dark matter searches for characterizing the subhalo mass function using galaxy-galaxy strong lenses~\citep{brehmer2019mining, 2022arXiv220509126A}, astrometric datasets~\citep{mishrasharma2022inferring}, stellar streams~\citep{hermans2021neuralsbi}, and halo clustering statistics~\citep{Dimitriou:2022cvc}. In this section, we will give a brief overview of the technique and summarize the salient aspects of our machine learning model. 

\subsection{Inference method}
\label{subsec:infmethod}

Suppose $\theta$ denotes the parameters of interest. Then, we can generate samples using a simulator (forward model) conditioned on $\theta$ that implicitly define the likelihood, $\{x\} \sim p(x \mid \theta)$. We can learn an estimator for the likelihood ratio between this distribution and one corresponding to an alternate reference hypothesis, $r(x\mid\theta) = \frac{p(x\mid\theta)}{p_{\mathrm{ref}}(x\mid\theta)}$, by training a classifier parameterized through $\theta$ to distinguish between samples drawn according to the two hypotheses. An ideal classifier would learn the decision function $s(x, \theta) = \frac{p(x\mid \theta)}{p(x\mid \theta) + p_{\mathrm{ref}}(x\mid\theta)}$ which is one-to-one with the aforementioned likelihood ratio~\citep{cranmer2015approximating,baldi2016parameterized,mohamed2016learning}:  
\begin{align}
     r(x\mid\theta) = \frac{s(x, \theta)}{1 - s(x, \theta)}.
     \label{eq:llr}
\end{align}
In neural likelihood-ratio estimation, the classifier $\hat s(x, \theta)$ is a neural network appropriate for the data modality in the given task and is trained through supervised learning. When the classifier decision is determined by applying a sigmoidal projection, the log-likelihood ratio estimates $\log \hat r(x\mid\theta)$ can be conveniently read off through the pre-sigmoidal outputs (logits) directly. 



In our specific inference task, $\theta$ is a single parameter that characterizes the power law slope $\gamma$ in the EPL profile and $x$ is a lensed image generated with an underlying $\gamma$. Having generated an ensemble of lensed image-parameter pairs, we train a classifier with a dependence on $\theta$ to distinguish between samples from the joint data-parameter distribution $p(x, \theta)$ and those from the reference distribution, $p_{\mathrm{ref}}(x, \theta)$, taken to be the product of the marginal distributions of the data and slope parameter, $\{x, \theta\}_{\mathrm{ref}} \sim p(x) p(\theta)$~\citep{hermans2019likelihoodfree}. Samples following the reference hypothesis are generated by shuffling the $\gamma$ labels in each batch so that they no longer align with the lensed images they generated. Intuitively, the classifier implicitly learns whether a $\gamma$ value is likely to be the true underlying slope that generated a given image by learning to distinguish between data-parameter pairs from the two distributions: the higher the classification score a pair receives, the more likely it is to have been generated according to $p(x, \theta)$. 

To compute the likelihood-ratio profile for a given lensed image as a function of our parameter of interest at test time, we obtain the classifier logits for a linearly-spaced array of input $\gamma$. The computational cost during test time therefore scales linearly with the number of test points in the parameter space. For our single parameter of interest $\gamma$, the inference is relatively efficient through direct computation; for a high-dimensional parameter, sampling techniques may be necessary~\citep{hermans2019likelihoodfree}. 

An aspect of likelihood-ratio estimation relevant to our application is that it allows us to conveniently combine information across multiple images with the same underlying $\gamma$ and obtain stronger constraints~\citep{brehmer2019mining}, as desired when working with an ensemble of strong lensing observations. If we assume that the individual inputs are independently and identically distributed when conditioned on the parameter of interest $\gamma$, then the combined likelihood ratio for a set of images $\{x\}$ can be obtained through the product of the individual likelihood ratios, 
\begin{align}
    \hat r(\{x\} \mid \gamma) = \prod_{i} \hat r(x_i \mid \gamma). 
    \label{eq:combine_r}
\end{align}
In contrast, when using simulation-based inference methods that directly learn a parameter posterior density on individual images, combining this information across images can be more challenging as it typically involves approximating integrals through sampling, which may introduce numerical inaccuracies that propagate when scaling to a large ensemble of lenses. 

\subsection{Model and training details}
\label{subsec:train_details}

We use a modified version of the ResNet-18 convolutional neural network implemented in \texttt{PyTorch}~\citep{NEURIPS2019_9015} as our classifier~\citep{he2016deepresiduallearning}. We output the classification score $\hat s(x,\theta)$ by projecting through a sigmoid activation after the dense layers of the ResNet. 
At training time, we append the true $\gamma$ to the latent vector after the convolutional layers, which is then passed into the dense layers. This ensures that the classifier output depends on both the image and the $\gamma$ value. Our training objective for the classification task is the canonical binary cross-entropy loss. 

The pixel values are normalized to zero mean and unit standard deviation across the training set. We also normalize the $\gamma$ values to zero mean. At test time, we use the training mean and standard deviation to whiten our test data so that the inputs into the model are consistent from training to testing. 

We performed a coarse grid hyperparameter search and chose an initial learning rate of $10^{-3}$ with the AdamW optimizer~\citep{kingma2014adam, loshchilove2017adamw} and batch size of 2000. We found that a larger batch size significantly increases the stability of training and produces better downstream results, consistent with previous findings~\citep{hermans2021neuralsbi, hoffer2017train}. During training, we follow a learning rate schedule that decays by an order of magnitude when the validation loss stagnates for 3 epochs, followed by a cool-down period of 2 epochs. We terminate our training when the validation loss plateaus for 6 epochs under a threshold of $10^{-3}$. Our training and validation sets consist of 50,000 and 1,000 mock images respectively, generated using the forward model described in Sec.~\ref{sec:data}. The model was trained on NVIDIA V100 GPUs with typical training time of 8 to 10 hours for 25 to 30 epochs.


\section{Results}
\label{sec:results}

\begin{figure*}
    \centering
    \includegraphics[width=\columnwidth]{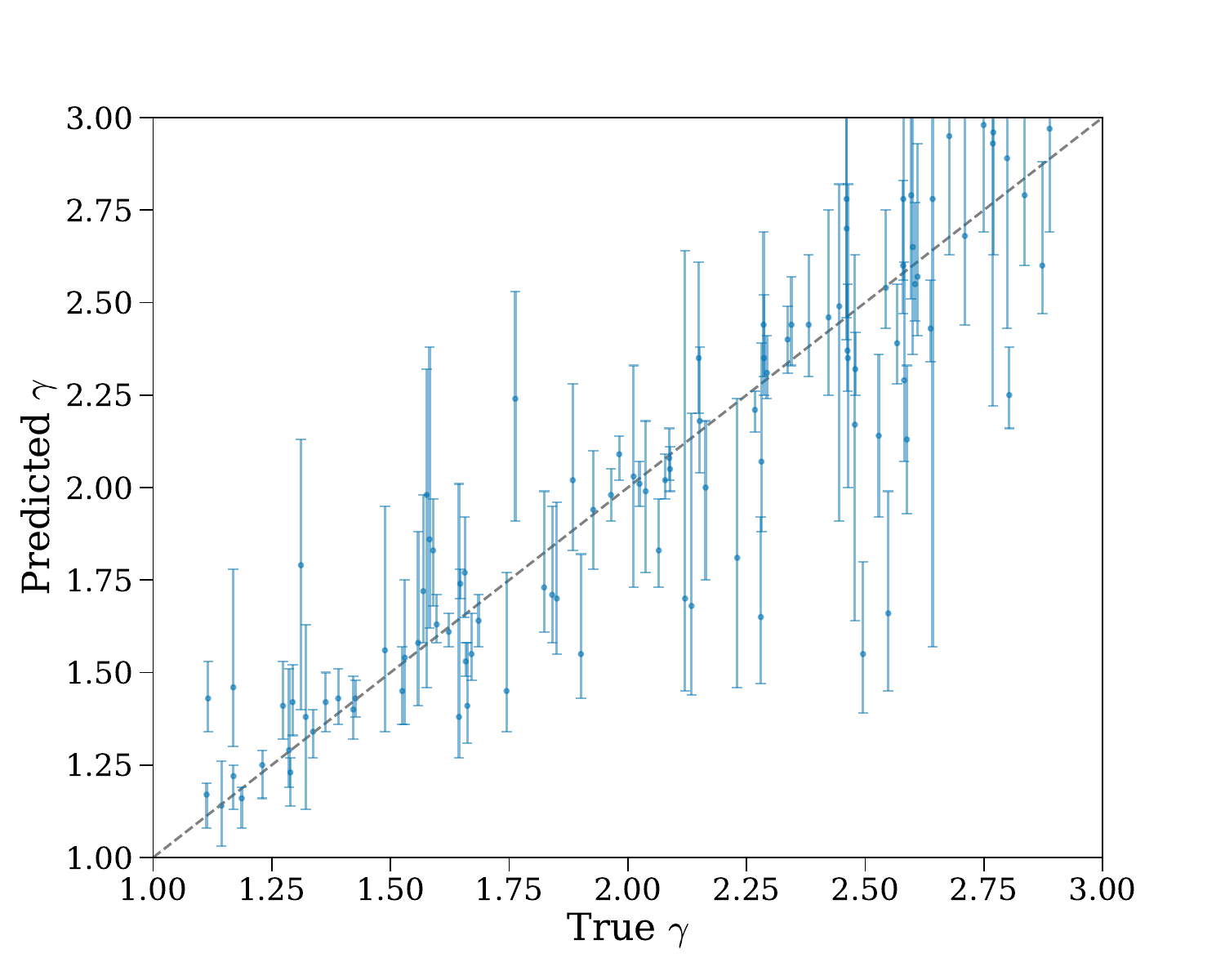}
    \includegraphics[width=\columnwidth]{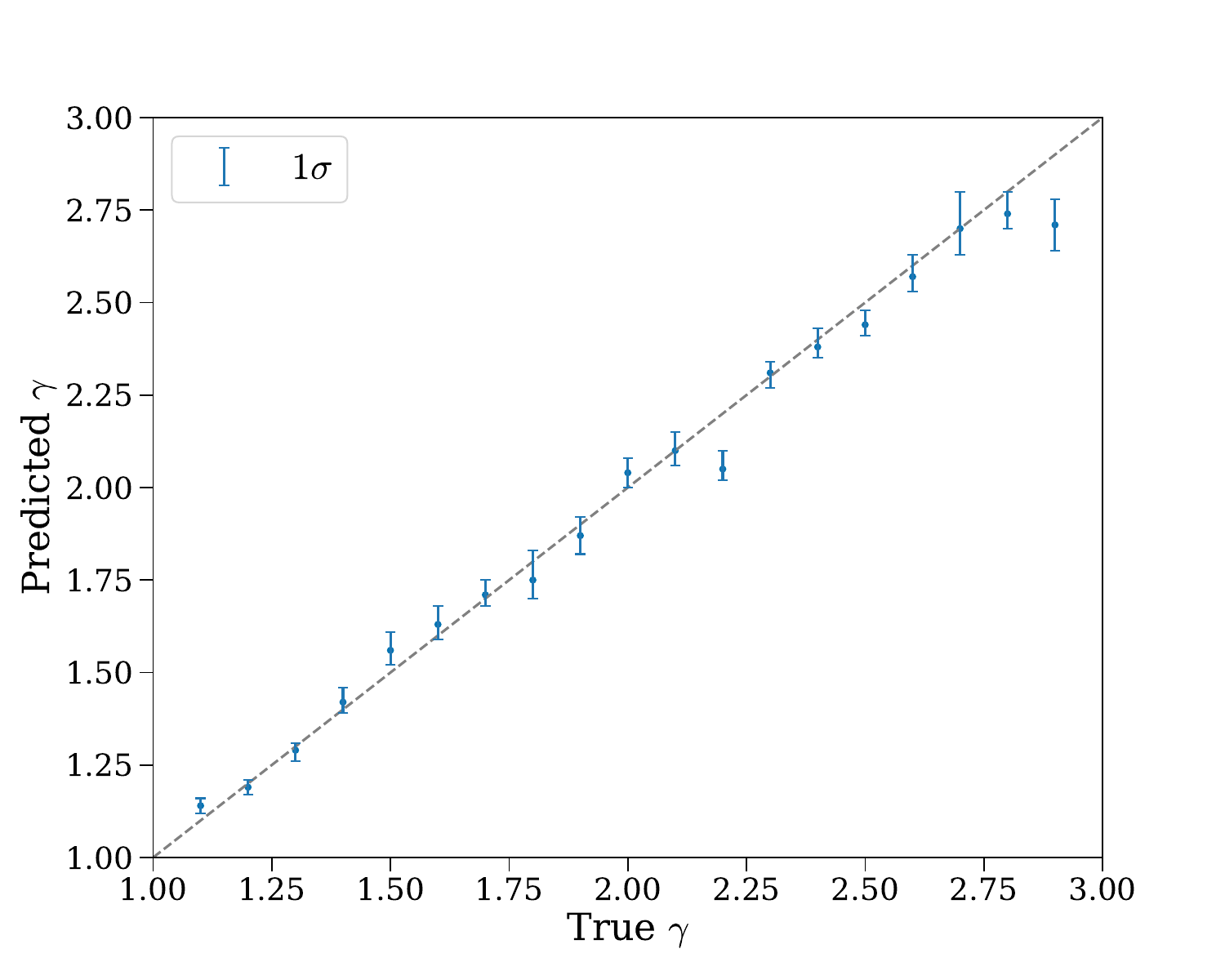}
    \caption{(\textit{Left}) Scatter plot of the maximum-likelihood $\gamma$ and associated 1-$\sigma$ uncertainties predicted using the trained likelihood-ratio estimator compared with the true underlying $\gamma$ of 100 test images with EPL subhalos. (\textit{Right}) Maximum-likelihood $\gamma$ and associated 1-$\sigma$ uncertainties predicted from combined likelihoods of sets of 10 images containing EPL subhalos compared with the true $\gamma$ underlying each set of images. For both panels, the model was trained on images containing EPL subhalos with $M_{200} \in [10^7, 10^{10}]\mathrm M_{\odot}$ and $N_{\mathrm{sub}} \sim \mathcal U\{0, 500\}$.}
    \label{fig:val_scatter}
\end{figure*}

To infer the slope of subhalos in a given image $x_i$, we feed the image along with an array of test $\gamma$ into the trained model to obtain the likelihood-ratio profile estimate $\hat{r}(x_i \mid \gamma)$. In Fig.~\ref{fig:val_scatter}, we show in the left panel the individual predictions compared to the true $\gamma$ for 200 test images containing EPL subhalos. We show the maximum-likelihood estimates $\hat{r}_\mathrm{MLE}(x_i \mid \gamma)$ using the scatter points and the $1\sigma$ uncertainties using error bars, assuming a $\chi^2$ distribution on the test-statistic $2 \left[\log  \hat{r}_\mathrm{MLE} - \log \hat{r}\right]$~\citep{Wilks:1938dza}.  Predictions can be seen to follow the true $\gamma$ values in trend, with reasonably large inferred uncertainties. 

To leverage the statistical effect of multiple observations, we can combine likelihood ratios from several images that share the same underlying $\gamma$, following Eq.~\eqref{eq:combine_r}. We show these results in the right panel of Fig.~\ref{fig:val_scatter}. Each prediction is obtained by combining the likelihood ratios (i.e., summing the log-likelihood ratios) of 50 images that share the same latent slope. As expected, combining information from a set of images significantly improves constraining power on the slope. Note that the predicted slope at the high-slope end has a small visible bias, likely due to the fact that we restrict the values of $\gamma$ to $[1.1, 2.9]$ so that there are relatively few images corresponding to $\gamma \gtrsim 2.9$ in the training set. However, we expect realistic images to have subhalos whose slopes lie reasonably far from the edges of the $(1, 3)$ range, so this bias has a negligible effect on our overall conclusions. 

\subsection{Region of maximum observability}
\label{subsec:rmo}

Our training set contains EPL subhalos because the EPL density profile allows us to conveniently label each image with its underlying slope. During test time, however, the EPL density profile is no longer sufficient for simulating realistic subhalos. With more realistic subhalo density profiles, the power-law slope is no longer constant at different radial distances from the center of a subhalo. For instance, the NFW profile scales as $r^{-1}$ near its core but asymptotically transitions to $r^{-3}$ past its scale radius. This necessitates a more concrete definition of ``slope.''

\citet{sengul2022probing} developed a formalism for estimating a region of maximum observability (RMO) and defined an effective slope for each perturber in a lensing system. We will sketch out the core steps in the RMO calculation and refer readers to the reference for a more detailed discussion. The central step in the RMO calculation is approximating the deviations in the brightness of each image pixel caused by a subhalo. This is done by performing a perturbative expansion in the deflection angles, assuming the deflections due to the subhalo are sufficiently small. This allows us to obtain a linear mapping between the pixel brightness deviations and the subhalo's average convergence. The RMO is then defined as the region in radial distance from the subhalo center in which the subhalo's average convergence measurement has the smallest uncertainty. The RMO depends on various parameters of a lensing system, particularly the gradient of the source light and the noise in each pixel, so it needs to be calculated for each system individually. When we constrain an effective power-law slope of a subhalo, the slope we measure is approximately the mean slope of the average convergence in the RMO following Eq.~\eqref{eq:slope_kappabar}. For the perturbative expansion in the RMO calculation to be valid, an important assumption is that the maximum angular deflection caused by the subhalo is much less than the pixel size of the image. In our case, the maximum angular deflection caused by an NFW subhalo with $M_{200} = 10^{10}~\mathrm M_{\odot}$ (which translates to $c_{200} \approx 10$) at $z_{\mathrm{lens}} = 0.5$ is around $0.009''$, which is much smaller than our pixel size of $0.08''$ and ensures the validity of our RMO approximation. 

\begin{figure}
    \centering
    \includegraphics[width=\columnwidth]{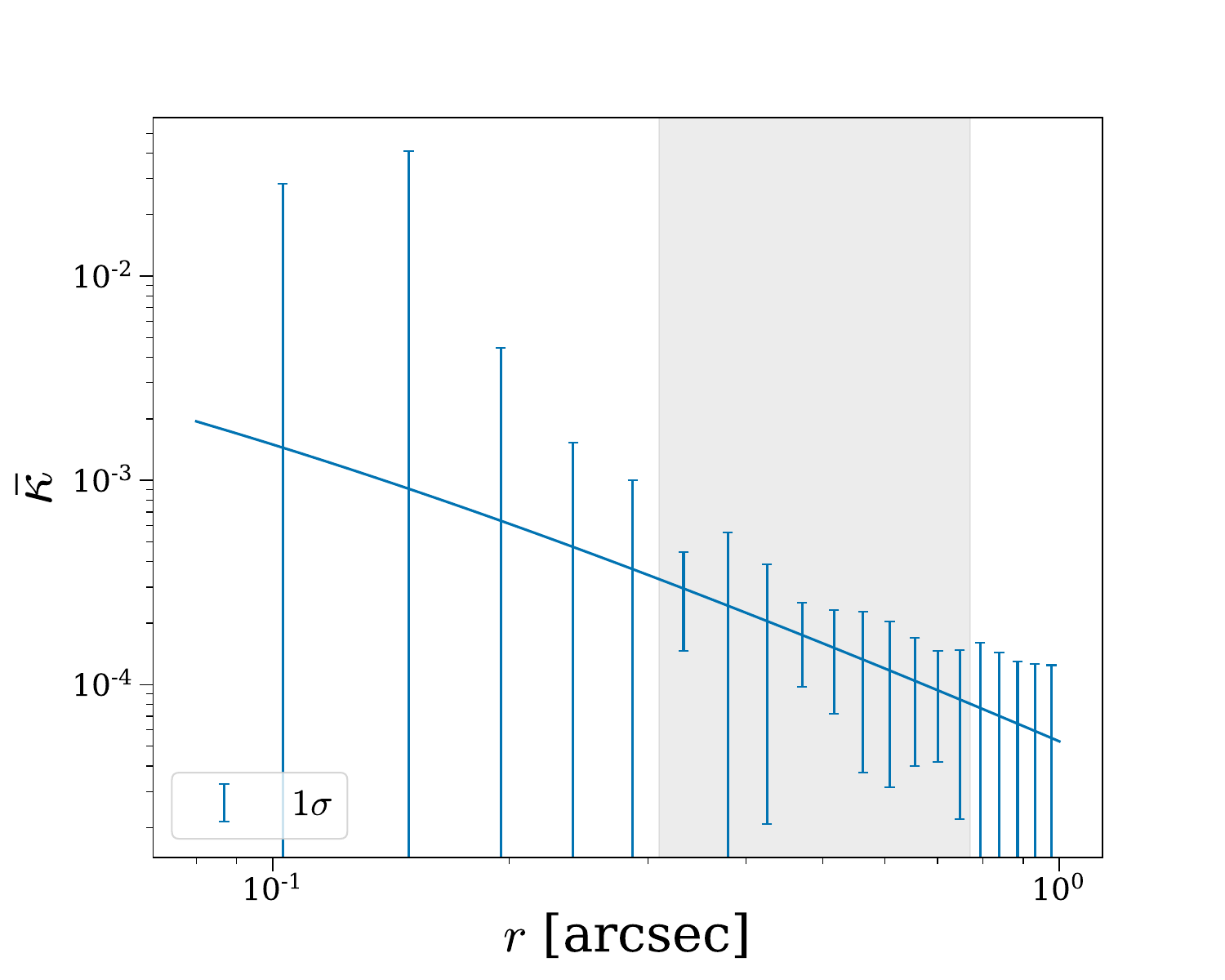}
    \includegraphics[width=\columnwidth]{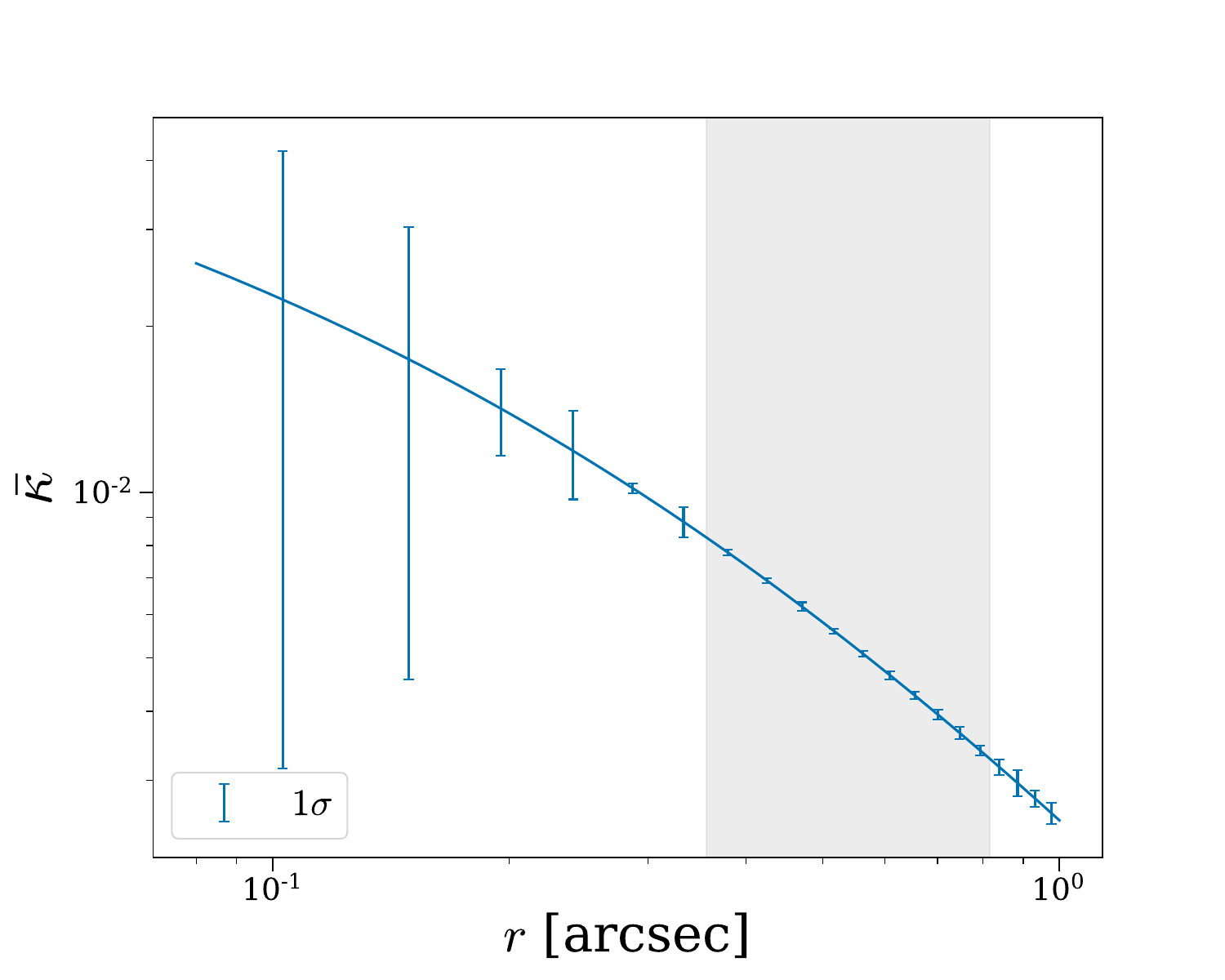}
    \caption{An example of the average convergence $\bar{\kappa}(r)$ within a radius $r$ from the subhalo center, for NFW subhalos with masses $M_{200} \approx 1.5 \times 10^{7}\mathrm M_{\odot}$ (\textit{top}) and $M_{200} \approx 1.6 \times 10^{9}\mathrm M_{\odot}$ (\textit{bottom}). The error bars show the $1\sigma$ errors on $\bar{\kappa}$ and the shaded gray area shows the estimated RMO where the relative errors on $\bar{\kappa}$ are less than twice the smallest relative error.}
    \label{fig:rmo_examples}
\end{figure}

Following \citet{sengul2022probing}, we define our RMO to be the region where the relative error on the average convergence is less than twice the smallest relative error. As an illustrative example, we take one of our test images and compute the RMO of two NFW subhalos of different masses. The two subhalos come from the same source-lens system but are located at different positions in the lens plane. In Fig.~\ref{fig:rmo_examples}, we show their average convergence $\bar{\kappa}$ as a function of radial distance $r$ in the lens plane (in blue curves) and their RMOs (in shaded gray). The upper panel shows the RMO of a subhalo with $M_{200} \approx 1.5 \times 10^{7}\mathrm M_{\odot}$, while the lower panel shows the RMO of a subhalo with $M_{200} \approx 1.6 \times 10^{9}\mathrm M_{\odot}$. The lower limit on the radial distance is set by the resolution of the image. As expected, the lower-mass subhalo has much larger errors on its average convergence profile than the higher-mass subhalo because low-mass subhalos have a low signal-to-noise ratio. As a consequence, the power-law slope of a low-mass subhalo is much less well-constrained than its high-mass counterpart. In both cases, when we take into account the uncertainties, the NFW subhalos are roughly indistinguishable from EPL subhalos. We compute the average power-law slope in the RMO for the two subhalos using Eq.~\eqref{eq:slope_kappabar}. The low-mass subhalo has an effective slope of approximately 2.55, while the high-mass subhalo has an effective slope of approximately 2.13. This agrees with the expectation that more massive subhalos have profiles that are farther extended radially and have flatter profiles than lighter subhalos when compared at the same radial distance. 

The RMO approximation allows us to obtain an estimate of the power-law slope for a single subhalo at a time. Because the masses of subhalos in each of our test images span three orders of magnitude, we obtain a different effective slope for each subhalo. In contrast, when we carry out inference with our likelihood-ratio estimator, we get a unique slope prediction for each image. Attaching a sensible physical interpretation to our model output is challenging because the neural network implicitly marginalizes over the information of all the subhalos in an image. Moreover, when we combine a set of likelihoods from multiple images, we further marginalize over source and main lens parameters, both of which affect the RMO approximation. It is nonetheless useful to compare the set of effective slopes computed analytically from the RMO with our model prediction to gain some physical understanding of the neural network output. 

\begin{figure}
    \centering
    \includegraphics[width=\columnwidth]{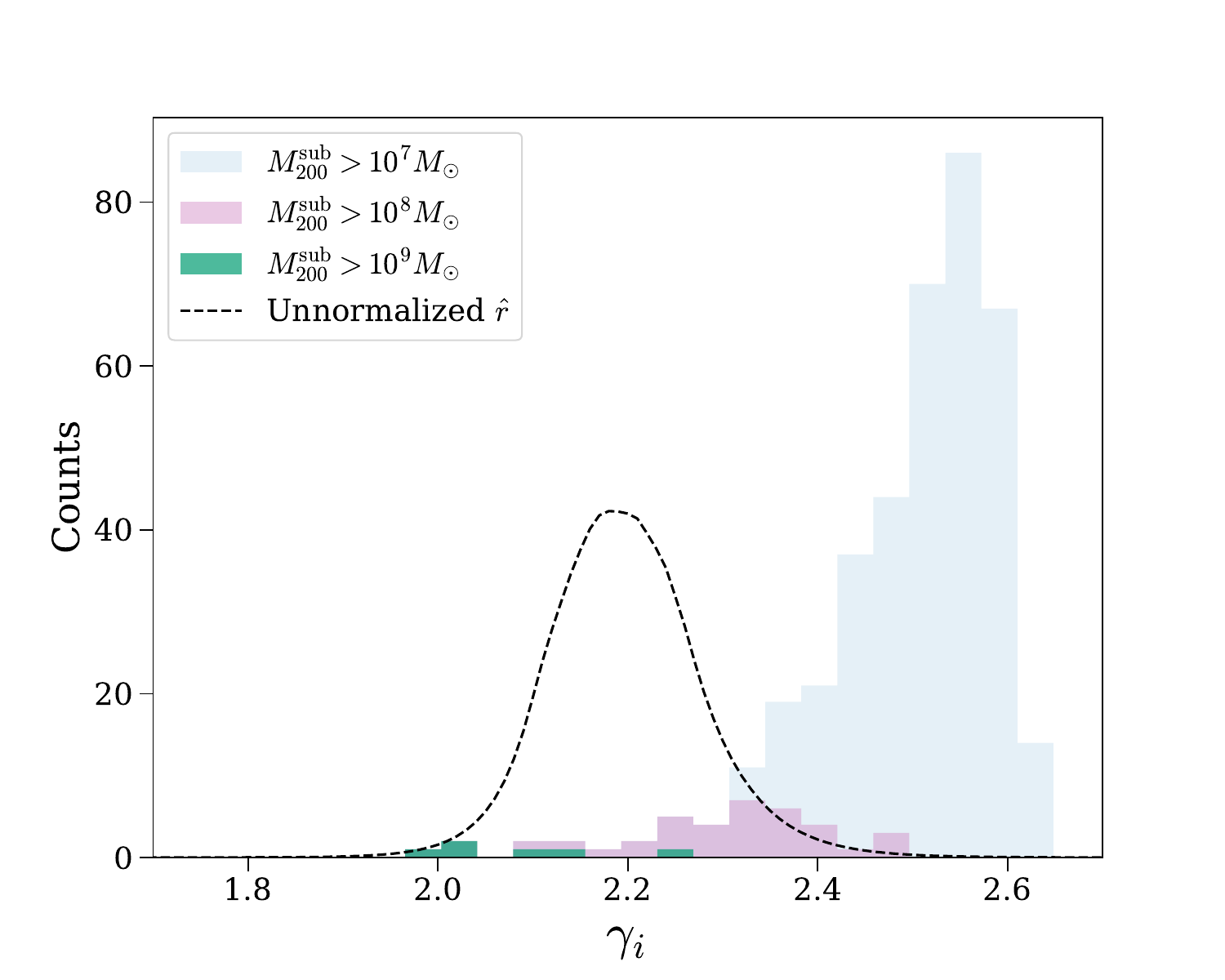}
    \caption{Histograms of the RMO effective power-law slope estimates of 388 NFW subhalos in a test set image separated by mass ranges of $[10^7, 10^{10}]\mathrm M_{\odot}$ (blue), $[10^8, 10^{10}]\mathrm M_{\odot}$ (magenta), and $[10^9, 10^{10}]\mathrm M_{\odot}$ (green). The likelihood-ratio estimate ($\hat{r}$) of the same image using the trained neural network model is scaled up for clarity and shown in black dashed line.}
    \label{fig:model_vs_rmoanalytical}
\end{figure}

To carry out this comparison, we focus on a single image and compare the RMO effective slopes of its subhalos with our model's prediction. In Fig.~\ref{fig:model_vs_rmoanalytical}, we show the set of analytically approximated RMO effective slopes for 388 NFW subhalos in an image, indicated by histograms corresponding to three mass ranges. These NFW subhalos all have concentrations given by $f_{\mathrm{CDM}}(M_{200})$, based on their individual masses. For visualization purposes, we then take the estimated likelihood ratio $\hat{r}$ of the same image using the trained neural likelihood-ratio estimator and scale it up so that it fits roughly on the same scale as the histograms. We show the likelihood-ratio estimate from our neural network in black dashed line. The RMO effective slopes at the low and high mass ranges in Fig.~\ref{fig:model_vs_rmoanalytical} both roughly agree with the examples shown in Fig.~\ref{fig:rmo_examples}. We see that $\hat{r}$ peaks at roughly the middle of the span of the histograms. It can be coarsely interpreted as a weighted average of the histogram where the more massive subhalos are given more weight than lighter subhalos. Intuitively, the more massive subhalos get more weight because subhalos in the $[10^7, 10^8]\mathrm M_{\odot}$ mass range have large errors in their RMOs so that their effective slope estimates also have large uncertainties. It may be tempting to conclude that the $\hat{r}$ likelihood is high-mass dominated and receives very little contribution from the low-mass subhalos. We will show later in Sec.~\ref{subsec:lowmass_vs_highmass} that this is in fact not the case. We will compare the $\hat{r}$ predictions from two different mass ranges and show that the model has significant sensitivity to the $[10^7, 10^8]\mathrm M_{\odot}$ subhalos.

\subsection{Test on NFW subhalos with varying concentrations}
\label{subsec:nfw_concentration}

As discussed in Sec.~\ref{subsec:rmo}, realistic subhalo profiles are close to being indistinguishable from EPL subhalos. Therefore, even though our likelihood-ratio estimator is trained on images containing EPL subhalos, we expect it to produce sensible predictions on more realistic subhalos at test time. 

To simulate subhalo populations with different slopes, we produce test sets with NFW subhalos while varying their mass-concentration relation, as discussed in Sec.~\ref{subsubsec:test_subhalo}. We expect that the predicted slopes of low-concentration multiplicative factor $a$ should be smaller than those of high multiplicative factor $a$ because the subhalo inner density profiles are flatter with lower concentrations. Because subhalos of different masses have different concentrations as well as different effective slopes, as a simplified toy experiment, we first work solely with $10^9\mathrm M_{\odot}$ subhalos. We train a model on mock images containing $N_{\mathrm{sub}} \sim \mathcal U\{0, 100\}$ subhalos of $M_{200} = 10^9\mathrm M_{\odot}$, and subsequently evaluate the model on test sets each with a different $c_{200}$. Note that the mass-concentration relation $f_{\mathrm{CDM}}(M_{200})$ gives $c_{200} = 12.2$ for a $M_{200} = 10^9\mathrm M_{\odot}$ subhalo. With these parameter choices, the total subhalo mass we add to the lensing systems is relatively large, so we add a negative mass sheet during modeling, as discussed in Sec.~\ref{subsubsec:test_subhalo}. 

\begin{figure}
    \centering
    \includegraphics[width=\columnwidth]{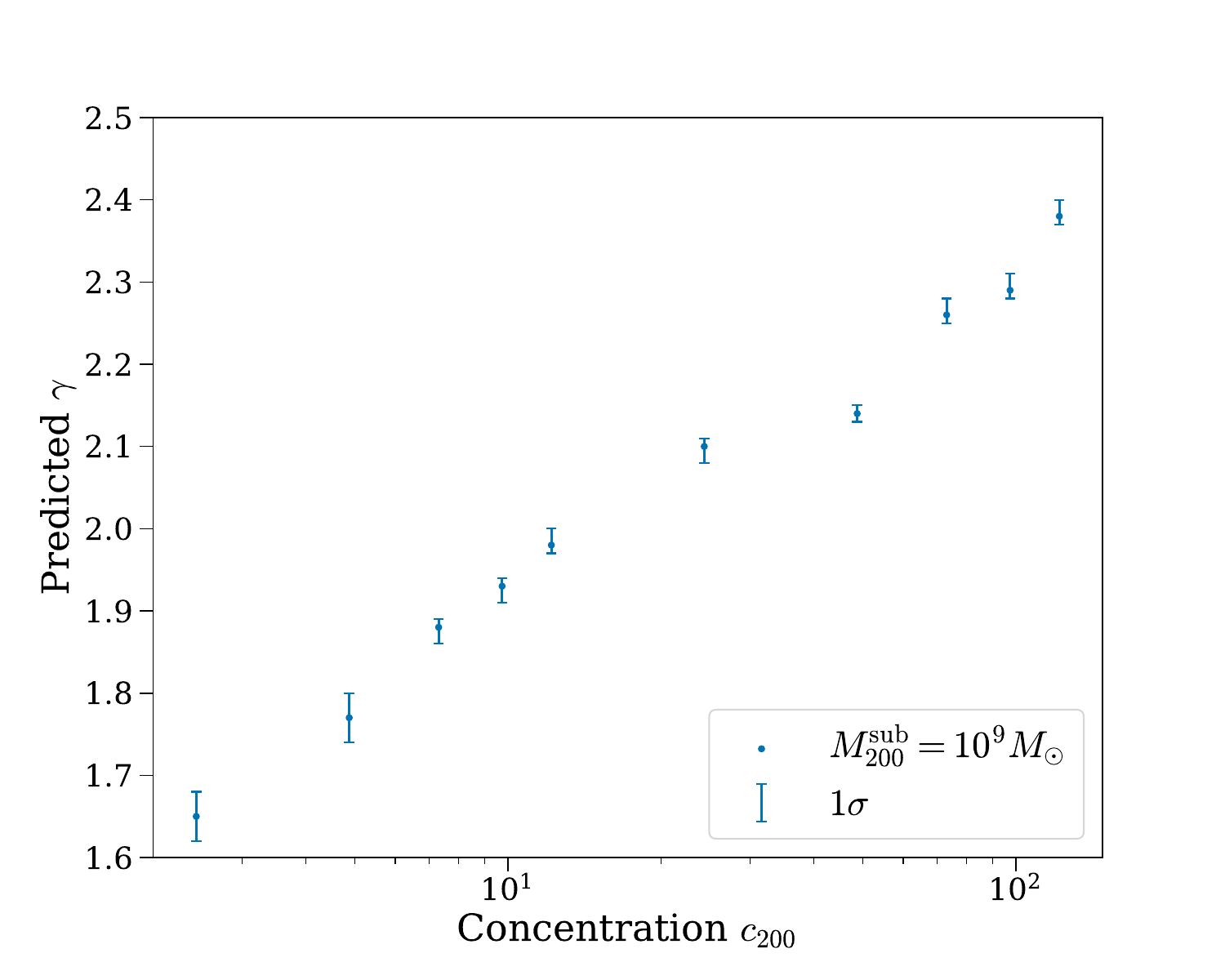}
    \caption{The maximum-likelihood predictions (scatter points) and $1\sigma$ uncertainties (error bars) obtained by combining 50 images with $N_{\mathrm{sub}} \sim \mathcal U\{0, 100\}$ of $10^9\mathrm M_{\odot}$ NFW subhalos as a function of the concentration $c_{200}$. The model used for inference was trained on images with $N_{\mathrm{sub}} \sim \mathcal U\{0, 100\}$ and $10^9\mathrm M_{\odot}$ EPL subhalos.}
    \label{fig:nfw_logm1e9}
\end{figure}

In Fig.~\ref{fig:nfw_logm1e9}, we show the model predictions by combining the likelihoods of 50 images in each test set. As expected, the model-predicted slope increases with the subhalo concentration. Remarkably, the model is able to effectively distinguish between different subhalo concentrations. Moreover, we see that the model predicts a slope of $\gamma \approx 2.1$ at approximately $c_{200} = 12.2$, which is consistent with the RMO slope estimate of the $1.6 \times 10^9\mathrm M_{\odot}$ subhalo in Fig.~\ref{fig:rmo_examples}.  

\subsection{Effect of low-mass subhalos}
\label{subsec:lowmass_vs_highmass}

One of the main reasons for harnessing population-level statistics of subhalos is that it allows us to extract information from the low-mass subhalos that are difficult, if not impossible, to be individually observed. It is therefore important to quantify the sensitivity of our model to low-mass subhalos. We can do so by comparing model predictions on two subhalo mass ranges: $[10^7, 10^{10}]\mathrm M_{\odot}$ and $[10^8, 10^{10}]\mathrm M_{\odot}$. The difference between the predictions on NFW subhalo populations in the two mass ranges will indicate the contributions from the low-mass subhalos. 

\begin{figure}
    \centering
    \includegraphics[width=\columnwidth]{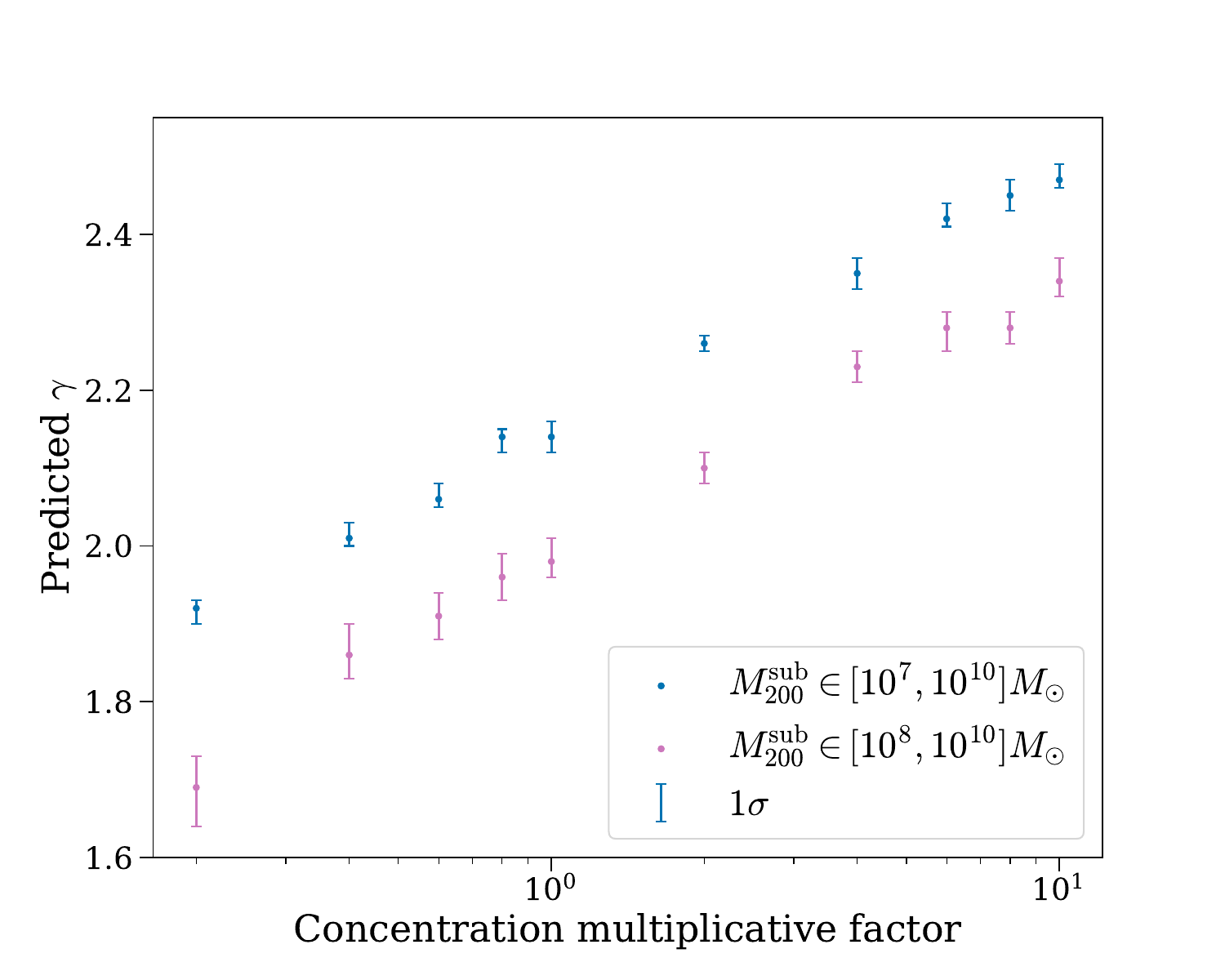}
    \caption{The maximum-likelihood predictions (scatter points) and $1\sigma$ uncertainties (error bars) obtained by combining 50 images with $N_{\mathrm{sub}} \sim \mathcal U\{0, 500\}$ and $M_{200} \in [10^7, 10^{10}]\mathrm M_{\odot}$ NFW subhalos as a function of the concentration $c_{200}$ (blue), compared to those with $N_{\mathrm{sub}} \sim \mathcal U\{0, 60\}$ and $M_{200} \in [10^8, 10^{10}]\mathrm M_{\odot}$ NFW subhalos (magenta).}
    \label{fig:lowmass_effect}
\end{figure}

To carry out this comparison, we generate new training and test sets with subhalo masses in the $[10^8, 10^{10}]\mathrm M_{\odot}$ range. These images are not meant to be realistic and, instead, serve the purpose of disentangling the effects of high- and low-mass subhalos on our model's slope predictions. In our $[10^7, 10^{10}]\mathrm M_{\odot}$ training set, we take $N_{\mathrm{sub}} \sim \mathcal U\{0, 500\}$, which translates to approximately $N_{\mathrm{sub}} \sim \mathcal U\{0, 60\}$ subhalos in the $[10^8, 10^{10}]\mathrm M_{\odot}$ range under the same assumed subhalo mass function. We keep all other lensing parameters the same between the two datasets. In Fig.~\ref{fig:lowmass_effect}, we show the maximum-likelihood model predictions of combining 50 images and their associated $1\sigma$ uncertainties on the low and high mass range datasets in blue and magenta, respectively. Because our datasets include subhalos of varying masses and thereby different $c_{200}$ values, we show the concentration multiplicative factor $a$ instead of $c_{200}$ in the $x$-axis. In the figure, we see a clear separation between the predictions on the two sets of images, indicating a significant contribution of the low-mass subhalos to the slope estimates at all concentration scales. This demonstrates that our model is sensitive to low-mass subhalos and incorporates their information when making predictions. Moreover, we notice that subhalos in the $[10^8, 10^{10}]\mathrm M_{\odot}$ mass range produce lower predicted slopes than their lower mass counterparts at all concentration factors. This is consistent with the lower slope predictions for more massive subhalos through the RMO formalism shown in Fig.~\ref{fig:rmo_examples} and Fig.~\ref{fig:model_vs_rmoanalytical}. As a robustness check, we also took our model trained on $[10^7, 10^{10}]\mathrm M_{\odot}$ population and tested it on the $[10^8, 10^{10}]\mathrm M_{\odot}$ population, and we observed a similar separation in the predicted slopes, as is shown in Fig.~\ref{fig:lowmass_effect}. This means that the model is sensitive to the low-mass subhalo contribution to the effective slopes regardless of the subhalo population during training. The $[10^8, 10^{10}]\mathrm M_{\odot}$ subhalo population can be thought of as having an extreme suppression in the low-mass region of the SHMF compared to the $[10^7, 10^{10}]\mathrm M_{\odot}$ population. Because our model is sensitive to this suppression, we expect it to be able to detect changes in the SHMF. Together with the results shown in Sec.~\ref{subsec:nfw_concentration}, we see that two factors affect the model's effective density slope prediction: collective changes in the subhalo density profiles and deviations in the SHMF. CDM and alternative DM models differ in their predictions of these two factors, so our model's ability to predict effective density slopes of subhalo populations has the potential to help discern different classes of DM models.

\section{Conclusions and outlook}
\label{sec:conlcusion}

Strong gravitational lensing provides a promising avenue for testing different classes of microphysical dark matter models on sub-galactic scales. Typically, previous works have focused on using individual subhalos or subhalo populations to constrain the subhalo mass function, a quantity that depends on the precise subhalo mass definition. An effective density slope of subhalos has been recently proposed as a reliable observable for probing dark matter properties. The advantage of measuring the effective subhalo density slope is that it makes no assumptions about its density profile, unlike the commonly used subhalo mass function. \citet{sengul2022probing} demonstrated that measuring the effective density slope can be achieved on the most massive subhalo in an individual lensing system. However, two challenges lie ahead of going beyond the most massive subhalo in a single lensing system: first, adding more subhalos increases the parameter space dimension significantly and makes sampling inefficient; second, modeling each lensing system may take $\mathcal{O}$(days) at a time, so it quickly becomes computationally unfeasible to analyze a large number of lensing systems together.

Recent advances in machine learning provide promising tools for overcoming these challenges. Machine learning methods are capable of performing inference by implicitly marginalizing over a large parameter space without the need to characterize the effect of individual subhalos. In this work, we have introduced a neural likelihood-ratio estimator to extract a population-level effective density slope. This method allows us to make use of additional information contained in the lensing forward model, including the effect of low-mass subhalos. In addition, when compared to other neural simulation-based inference techniques such as those based on posterior density estimation, neural likelihood-ratio estimation can be used to more efficiently combine information across a large dataset.

Through testing our model on mock strong lensing images, we showed that the inferred effective density slope incorporates significant information from the low-mass subhalos and further demonstrated that our model is capable of distinguishing between subhalo populations with different density slopes. Different DM models predict effective density slopes of subhalo populations that deviate from the CDM prediction. For example, warm dark matter models predict a suppression in the number of low-mass subhalos and more cored subhalo density profiles~\citep{bode2001wdm, lovell2012wdm, lovell2014wdm}, and both of these effects will result in a smaller subhalo effective density slope prediction compared to that of the CDM model. Self-interacting dark matter models also predict more cored subhalo density profiles and thus will also produce effective density slope predictions smaller than that of the CDM model~\citep{spergel2000sidm, vogelsberger2012sidm, rocha2013sidm, kahlhoefer2019sidm}. In addition, certain self-interacting dark matter models predict core-collapsed subhalos that have particularly steep density slopes~\citep[e.g.][]{lyndenbell1968, 2000kochanek, 2002colin}. This means that our likelihood-ratio estimator's ability to distinguish between subhalo populations with different effective density slopes has significant implications for discerning DM scenarios. 

Several improvements to our analysis pipeline are necessary before being able to deploy it on real lensing data.
First, our image modeling procedure does not include line-of-sight halos that also affect the lensing signature. Second, because of the need for well-resolved source galaxies, the COSMOS galaxies used by \textsc{paltas} have original redshifts lower than the $z_{\mathrm{source}} = 1$ used in modeling our lensing systems. There are differences in the morphology of galaxies at different redshifts, so to simulate more realistic images, galaxies with better resolutions and at higher redshifts could be used in the future. Third, in our mock images, we assume that the lens light has been subtracted, but to more realistically model the lensed systems we could include the lens light using real galaxy images in the forward model as done for the source galaxy. Finally, we demonstrated our model's sensitivity to the subhalo effective density slopes by testing on NFW subhalos with varying concentrations; in a realistic application, it would be necessary to understand the expected effect of different microphysical DM models as well. Despite these additional complexities, we are hopeful that the neural likelihood-ratio estimator will be a useful tool for characterizing the inner slopes of dark matter subhalos using the influx of strong lensing systems that will be discovered by upcoming imaging surveys.

\section*{Acknowledgements}
We would like to thank Douglas Finkbeiner, Çağan \c{S}eng\"ul and Arthur Tsang for valuable discussions. CD and GZ are partially supported by NSF grant AST-1813694.
SM is partially supported by the U.S. Department of Energy, Office of Science, Office of High Energy Physics of U.S. Department of Energy under grant Contract Number DE-SC0012567.
This work is supported by the National Science Foundation under Cooperative Agreement PHY-2019786 (The NSF AI Institute for Artificial Intelligence and Fundamental Interactions, \url{http://iaifi.org/}).
The computations in this paper were run on the FASRC Cannon cluster supported by the FAS Division of Science Research Computing Group at Harvard University.

\section*{Data Availability}
The code used to produce the results shown in this paper is publicly available at \url{https://github.com/gemyxzhang/neural-subhalo-slope}. 

\bibliography{references} 

\begin{thebibliography}{}
\makeatletter
\relax
\def\mn@urlcharsother{\let\do\@makeother \do\$\do\&\do\#\do\^\do\_\do\%\do\~}
\def\mn@doi{\begingroup\mn@urlcharsother \@ifnextchar [ {\mn@doi@}
  {\mn@doi@[]}}
\def\mn@doi@[#1]#2{\def\@tempa{#1}\ifx\@tempa\@empty \href
  {http://dx.doi.org/#2} {doi:#2}\else \href {http://dx.doi.org/#2} {#1}\fi
  \endgroup}
\def\mn@eprint#1#2{\mn@eprint@#1:#2::\@nil}
\def\mn@eprint@arXiv#1{\href {http://arxiv.org/abs/#1} {{\tt arXiv:#1}}}
\def\mn@eprint@dblp#1{\href {http://dblp.uni-trier.de/rec/bibtex/#1.xml}
  {dblp:#1}}
\def\mn@eprint@#1:#2:#3:#4\@nil{\def\@tempa {#1}\def\@tempb {#2}\def\@tempc
  {#3}\ifx \@tempc \@empty \let \@tempc \@tempb \let \@tempb \@tempa \fi \ifx
  \@tempb \@empty \def\@tempb {arXiv}\fi \@ifundefined
  {mn@eprint@\@tempb}{\@tempb:\@tempc}{\expandafter \expandafter \csname
  mn@eprint@\@tempb\endcsname \expandafter{\@tempc}}}

\bibitem[\protect\citeauthoryear{Alsing, Wandelt  \& Feeney}{Alsing
  et~al.}{2018}]{delfi1}
Alsing J.,  Wandelt B.,   Feeney S.,  2018, \mn@doi [\mnras]
  {10.1093/mnras/sty819}, 477, 2874–2885

\bibitem[\protect\citeauthoryear{Alsing, Charnock, Feeney  \& Wandelt}{Alsing
  et~al.}{2019}]{Alsing2019fastlldfree}
Alsing J.,  Charnock T.,  Feeney S.,   Wandelt B.,  2019, \mn@doi [\mnras]
  {10.1093/mnras/stz1960}, 488, 4440

\bibitem[\protect\citeauthoryear{{Amorisco} et~al.,}{{Amorisco}
  et~al.}{2022}]{amorisco2022}
{Amorisco} N.~C.,  et~al., 2022, \mn@doi [\mnras] {10.1093/mnras/stab3527},
  \href {https://ui.adsabs.harvard.edu/abs/2022MNRAS.510.2464A} {510, 2464}

\bibitem[\protect\citeauthoryear{{Anau Montel}, {Coogan}, {Correa}, {Karchev}
  \& {Weniger}}{{Anau Montel} et~al.}{2022}]{2022arXiv220509126A}
{Anau Montel} N.,  {Coogan} A.,  {Correa} C.,  {Karchev} K.,   {Weniger} C.,
  2022, arXiv e-prints, \href
  {https://ui.adsabs.harvard.edu/abs/2022arXiv220509126A} {p. arXiv:2205.09126}

\bibitem[\protect\citeauthoryear{{Anderhalden}, {Schneider}, {Macci{\`o}},
  {Diemand}  \& {Bertone}}{{Anderhalden} et~al.}{2013}]{anderhalden2013hints}
{Anderhalden} D.,  {Schneider} A.,  {Macci{\`o}} A.~V.,  {Diemand} J.,
  {Bertone} G.,  2013, \mn@doi [\jcap] {10.1088/1475-7516/2013/03/014}, \href
  {https://ui.adsabs.harvard.edu/abs/2013JCAP...03..014A} {2013, 014}

\bibitem[\protect\citeauthoryear{{Baldi}, {Cranmer}, {Faucett}, {Sadowski}  \&
  {Whiteson}}{{Baldi} et~al.}{2016}]{baldi2016parameterized}
{Baldi} P.,  {Cranmer} K.,  {Faucett} T.,  {Sadowski} P.,   {Whiteson} D.,
  2016, \mn@doi [European Physical Journal C] {10.1140/epjc/s10052-016-4099-4},
  \href {https://ui.adsabs.harvard.edu/abs/2016EPJC...76..235B} {76, 235}

\bibitem[\protect\citeauthoryear{{Banik}, {Bovy}, {Bertone}, {Erkal}  \& {de
  Boer}}{{Banik} et~al.}{2021}]{banik2021novel}
{Banik} N.,  {Bovy} J.,  {Bertone} G.,  {Erkal} D.,   {de Boer} T.~J.~L.,
  2021, \mn@doi [\jcap] {10.1088/1475-7516/2021/10/043}, \href
  {https://ui.adsabs.harvard.edu/abs/2021JCAP...10..043B} {2021, 043}

\bibitem[\protect\citeauthoryear{{Barkana}}{{Barkana}}{1998}]{barkana1998epl}
{Barkana} R.,  1998, \mn@doi [\apj] {10.1086/305950}, \href
  {https://ui.adsabs.harvard.edu/abs/1998ApJ...502..531B} {502, 531}

\bibitem[\protect\citeauthoryear{{Bechtol} et~al.,}{{Bechtol}
  et~al.}{2019}]{bechtol2019lsst}
{Bechtol} K.,  et~al., 2019, \baas, \href
  {https://ui.adsabs.harvard.edu/abs/2019BAAS...51c.207B} {51, 207}

\bibitem[\protect\citeauthoryear{{Birrer} \& {Amara}}{{Birrer} \&
  {Amara}}{2018}]{birrer2018lenstronomy}
{Birrer} S.,  {Amara} A.,  2018, \mn@doi [Physics of the Dark Universe]
  {10.1016/j.dark.2018.11.002}, \href
  {https://ui.adsabs.harvard.edu/abs/2018PDU....22..189B} {22, 189}

\bibitem[\protect\citeauthoryear{{Birrer}, {Amara}  \& {Refregier}}{{Birrer}
  et~al.}{2015}]{birrer2015gravitational}
{Birrer} S.,  {Amara} A.,   {Refregier} A.,  2015, \mn@doi [\apj]
  {10.1088/0004-637X/813/2/102}, \href
  {https://ui.adsabs.harvard.edu/abs/2015ApJ...813..102B} {813, 102}

\bibitem[\protect\citeauthoryear{Birrer, Amara  \& Refregier}{Birrer
  et~al.}{2017}]{birrer2017lensingsubs}
Birrer S.,  Amara A.,   Refregier A.,  2017, \mn@doi [JCAP]
  {10.1088/1475-7516/2017/05/037}, 05, 037

\bibitem[\protect\citeauthoryear{{Bode}, {Ostriker}  \& {Turok}}{{Bode}
  et~al.}{2001}]{bode2001wdm}
{Bode} P.,  {Ostriker} J.~P.,   {Turok} N.,  2001, \mn@doi [\apj]
  {10.1086/321541}, \href
  {https://ui.adsabs.harvard.edu/abs/2001ApJ...556...93B} {556, 93}

\bibitem[\protect\citeauthoryear{{Bolton}, {Burles}, {Koopmans}, {Treu},
  {Gavazzi}, {Moustakas}, {Wayth}  \& {Schlegel}}{{Bolton}
  et~al.}{2008}]{bolton2008slacs}
{Bolton} A.~S.,  {Burles} S.,  {Koopmans} L. V.~E.,  {Treu} T.,  {Gavazzi} R.,
  {Moustakas} L.~A.,  {Wayth} R.,   {Schlegel} D.~J.,  2008, \mn@doi [\apj]
  {10.1086/589327}, \href
  {https://ui.adsabs.harvard.edu/abs/2008ApJ...682..964B} {682, 964}

\bibitem[\protect\citeauthoryear{{Bonaca} et~al.,}{{Bonaca}
  et~al.}{2020}]{bonaca2020highres}
{Bonaca} A.,  et~al., 2020, \mn@doi [\apjl] {10.3847/2041-8213/ab800c}, \href
  {https://ui.adsabs.harvard.edu/abs/2020ApJ...892L..37B} {892, L37}

\bibitem[\protect\citeauthoryear{{Brehmer}, {Mishra-Sharma}, {Hermans},
  {Louppe}  \& {Cranmer}}{{Brehmer} et~al.}{2019}]{brehmer2019mining}
{Brehmer} J.,  {Mishra-Sharma} S.,  {Hermans} J.,  {Louppe} G.,   {Cranmer} K.,
   2019, \mn@doi [\apj] {10.3847/1538-4357/ab4c41}, \href
  {https://ui.adsabs.harvard.edu/abs/2019ApJ...886...49B} {886, 49}

\bibitem[\protect\citeauthoryear{Brennan, Benson, Cyr-Racine, Keeton, Moustakas
   \& Pullen}{Brennan et~al.}{2019}]{brennan2018quantifying}
Brennan S.,  Benson A.~J.,  Cyr-Racine F.-Y.,  Keeton C.~R.,  Moustakas L.~A.,
   Pullen A.~R.,  2019, \mn@doi [Mon. Not. Roy. Astron. Soc.]
  {10.1093/mnras/stz1607}, 488, 5085

\bibitem[\protect\citeauthoryear{{Brewer}, {Huijser}  \& {Lewis}}{{Brewer}
  et~al.}{2016}]{brewer2016transdim}
{Brewer} B.~J.,  {Huijser} D.,   {Lewis} G.~F.,  2016, \mn@doi [\mnras]
  {10.1093/mnras/stv2370}, \href
  {https://ui.adsabs.harvard.edu/abs/2016MNRAS.455.1819B} {455, 1819}

\bibitem[\protect\citeauthoryear{{Brownstein} et~al.,}{{Brownstein}
  et~al.}{2012}]{brownstein201bells}
{Brownstein} J.~R.,  et~al., 2012, \mn@doi [\apj] {10.1088/0004-637X/744/1/41},
  \href {https://ui.adsabs.harvard.edu/abs/2012ApJ...744...41B} {744, 41}

\bibitem[\protect\citeauthoryear{{Buckley} \& {Peter}}{{Buckley} \&
  {Peter}}{2018}]{buckley2018gravitational}
{Buckley} M.~R.,  {Peter} A. H.~G.,  2018, \mn@doi [\physrep]
  {10.1016/j.physrep.2018.07.003}, \href
  {https://ui.adsabs.harvard.edu/abs/2018PhR...761....1B} {761, 1}

\bibitem[\protect\citeauthoryear{{Bullock} \& {Boylan-Kolchin}}{{Bullock} \&
  {Boylan-Kolchin}}{2017}]{bullock2017smallscale}
{Bullock} J.~S.,  {Boylan-Kolchin} M.,  2017, \mn@doi [\araa]
  {10.1146/annurev-astro-091916-055313}, \href
  {https://ui.adsabs.harvard.edu/abs/2017ARA&A..55..343B} {55, 343}

\bibitem[\protect\citeauthoryear{{Buschmann}, {Kopp}, {Safdi}  \&
  {Wu}}{{Buschmann} et~al.}{2018}]{buschmann2018stellar}
{Buschmann} M.,  {Kopp} J.,  {Safdi} B.~R.,   {Wu} C.-L.,  2018, \mn@doi [\prl]
  {10.1103/PhysRevLett.120.211101}, \href
  {https://ui.adsabs.harvard.edu/abs/2018PhRvL.120u1101B} {120, 211101}

\bibitem[\protect\citeauthoryear{{Carlberg}}{{Carlberg}}{2016}]{carlberg2016modeling}
{Carlberg} R.~G.,  2016, \mn@doi [\apj] {10.3847/0004-637X/820/1/45}, \href
  {https://ui.adsabs.harvard.edu/abs/2016ApJ...820...45C} {820, 45}

\bibitem[\protect\citeauthoryear{{Cole}, {Miller}, {Witte}, {Cai}, {Grootes},
  {Nattino}  \& {Weniger}}{{Cole} et~al.}{2022}]{Cole:2021gwr}
{Cole} A.,  {Miller} B.~K.,  {Witte} S.~J.,  {Cai} M.~X.,  {Grootes} M.~W.,
  {Nattino} F.,   {Weniger} C.,  2022, \mn@doi [\jcap]
  {10.1088/1475-7516/2022/09/004}, \href
  {https://ui.adsabs.harvard.edu/abs/2022JCAP...09..004C} {2022, 004}

\bibitem[\protect\citeauthoryear{{Col{\'\i}n}, {Avila-Reese}, {Valenzuela}  \&
  {Firmani}}{{Col{\'\i}n} et~al.}{2002}]{2002colin}
{Col{\'\i}n} P.,  {Avila-Reese} V.,  {Valenzuela} O.,   {Firmani} C.,  2002,
  \mn@doi [\apj] {10.1086/344259}, \href
  {https://ui.adsabs.harvard.edu/abs/2002ApJ...581..777C} {581, 777}

\bibitem[\protect\citeauthoryear{Collett}{Collett}{2015}]{Collett:2015roa}
Collett T.~E.,  2015, \mn@doi [Astrophys. J.] {10.1088/0004-637X/811/1/20},
  811, 20

\bibitem[\protect\citeauthoryear{Coogan, Karchev  \& Weniger}{Coogan
  et~al.}{2020}]{Coogan2020targeted}
Coogan A.,  Karchev K.,   Weniger C.,  2020, in {34th Conference on Neural
  Information Processing Systems}.  (\mn@eprint {arXiv} {2010.07032})

\bibitem[\protect\citeauthoryear{{Cranmer}, {Pavez}  \& {Louppe}}{{Cranmer}
  et~al.}{2015}]{cranmer2015approximating}
{Cranmer} K.,  {Pavez} J.,   {Louppe} G.,  2015, arXiv e-prints, \href
  {https://ui.adsabs.harvard.edu/abs/2015arXiv150602169C} {p. arXiv:1506.02169}

\bibitem[\protect\citeauthoryear{Cyr-Racine, Moustakas, Keeton, Sigurdson  \&
  Gilman}{Cyr-Racine et~al.}{2016}]{cyr-Racine2015darkcensus}
Cyr-Racine F.-Y.,  Moustakas L.~A.,  Keeton C.~R.,  Sigurdson K.,   Gilman
  D.~A.,  2016, \mn@doi [Phys. Rev. D] {10.1103/PhysRevD.94.043505}, 94, 043505

\bibitem[\protect\citeauthoryear{{Cyr-Racine}, {Keeton}  \&
  {Moustakas}}{{Cyr-Racine} et~al.}{2019}]{cyrracine2019}
{Cyr-Racine} F.-Y.,  {Keeton} C.~R.,   {Moustakas} L.~A.,  2019, \mn@doi [\prd]
  {10.1103/PhysRevD.100.023013}, \href
  {https://ui.adsabs.harvard.edu/abs/2019PhRvD.100b3013C} {100, 023013}

\bibitem[\protect\citeauthoryear{{Dai} \& {Seljak}}{{Dai} \&
  {Seljak}}{2022}]{Dai2022trenf}
{Dai} B.,  {Seljak} U.,  2022, \mn@doi [\mnras] {10.1093/mnras/stac2010}, \href
  {https://ui.adsabs.harvard.edu/abs/2022MNRAS.516.2363D} {516, 2363}

\bibitem[\protect\citeauthoryear{Dalal \& Kochanek}{Dalal \&
  Kochanek}{2002}]{dalal2001direction}
Dalal N.,  Kochanek C.~S.,  2002, \mn@doi [Astrophys. J.] {10.1086/340303},
  572, 25

\bibitem[\protect\citeauthoryear{{Daylan}, {Cyr-Racine}, {Diaz Rivero},
  {Dvorkin}  \& {Finkbeiner}}{{Daylan} et~al.}{2018}]{daylan2018probing}
{Daylan} T.,  {Cyr-Racine} F.-Y.,  {Diaz Rivero} A.,  {Dvorkin} C.,
  {Finkbeiner} D.~P.,  2018, \mn@doi [\apj] {10.3847/1538-4357/aaaa1e}, \href
  {https://ui.adsabs.harvard.edu/abs/2018ApJ...854..141D} {854, 141}

\bibitem[\protect\citeauthoryear{D\'\i{}az~Rivero, Cyr-Racine  \&
  Dvorkin}{D\'\i{}az~Rivero et~al.}{2018a}]{diazRivero2017powerspec}
D\'\i{}az~Rivero A.,  Cyr-Racine F.-Y.,   Dvorkin C.,  2018a, \mn@doi [Phys.
  Rev. D] {10.1103/PhysRevD.97.023001}, 97, 023001

\bibitem[\protect\citeauthoryear{D\'\i{}az~Rivero, Dvorkin, Cyr-Racine, Zavala
  \& Vogelsberger}{D\'\i{}az~Rivero
  et~al.}{2018b}]{diazRivero2018powerspecethos}
D\'\i{}az~Rivero A.,  Dvorkin C.,  Cyr-Racine F.-Y.,  Zavala J.,   Vogelsberger
  M.,  2018b, \mn@doi [Phys. Rev. D] {10.1103/PhysRevD.98.103517}, 98, 103517

\bibitem[\protect\citeauthoryear{Dimitriou, Weniger  \& Correa}{Dimitriou
  et~al.}{2022}]{Dimitriou:2022cvc}
Dimitriou A.,  Weniger C.,   Correa C.~A.,  2022, arXiv e-prints

\bibitem[\protect\citeauthoryear{{Dutton} \& {Macci{\`o}}}{{Dutton} \&
  {Macci{\`o}}}{2014}]{dutton2014cdmhalo}
{Dutton} A.~A.,  {Macci{\`o}} A.~V.,  2014, \mn@doi [\mnras]
  {10.1093/mnras/stu742}, \href
  {https://ui.adsabs.harvard.edu/abs/2014MNRAS.441.3359D} {441, 3359}

\bibitem[\protect\citeauthoryear{{Fadely} \& {Keeton}}{{Fadely} \&
  {Keeton}}{2012}]{fadely2012substructure}
{Fadely} R.,  {Keeton} C.~R.,  2012, \mn@doi [\mnras]
  {10.1111/j.1365-2966.2011.19729.x}, \href
  {https://ui.adsabs.harvard.edu/abs/2012MNRAS.419..936F} {419, 936}

\bibitem[\protect\citeauthoryear{{Feldmann} \& {Spolyar}}{{Feldmann} \&
  {Spolyar}}{2015}]{feldmann2015detecting}
{Feldmann} R.,  {Spolyar} D.,  2015, \mn@doi [\mnras] {10.1093/mnras/stu2147},
  \href {https://ui.adsabs.harvard.edu/abs/2015MNRAS.446.1000F} {446, 1000}

\bibitem[\protect\citeauthoryear{{Fitts} et~al.,}{{Fitts}
  et~al.}{2017}]{fitts2017fire}
{Fitts} A.,  et~al., 2017, \mn@doi [\mnras] {10.1093/mnras/stx1757}, \href
  {https://ui.adsabs.harvard.edu/abs/2017MNRAS.471.3547F} {471, 3547}

\bibitem[\protect\citeauthoryear{{Gilman}, {Birrer}, {Treu}, {Keeton}  \&
  {Nierenberg}}{{Gilman} et~al.}{2018}]{gilman2018}
{Gilman} D.,  {Birrer} S.,  {Treu} T.,  {Keeton} C.~R.,   {Nierenberg} A.,
  2018, \mn@doi [\mnras] {10.1093/mnras/sty2261}, \href
  {https://ui.adsabs.harvard.edu/abs/2018MNRAS.481..819G} {481, 819}

\bibitem[\protect\citeauthoryear{{Gilman}, {Birrer}, {Nierenberg}, {Treu}, {Du}
   \& {Benson}}{{Gilman} et~al.}{2020}]{gilman2020wdm}
{Gilman} D.,  {Birrer} S.,  {Nierenberg} A.,  {Treu} T.,  {Du} X.,   {Benson}
  A.,  2020, \mn@doi [\mnras] {10.1093/mnras/stz3480}, \href
  {https://ui.adsabs.harvard.edu/abs/2020MNRAS.491.6077G} {491, 6077}

\bibitem[\protect\citeauthoryear{{Gilman}, {Bovy}, {Treu}, {Nierenberg},
  {Birrer}, {Benson}  \& {Sameie}}{{Gilman} et~al.}{2021}]{gilman2021strong}
{Gilman} D.,  {Bovy} J.,  {Treu} T.,  {Nierenberg} A.,  {Birrer} S.,  {Benson}
  A.,   {Sameie} O.,  2021, \mn@doi [\mnras] {10.1093/mnras/stab2335}, \href
  {https://ui.adsabs.harvard.edu/abs/2021MNRAS.507.2432G} {507, 2432}

\bibitem[\protect\citeauthoryear{He, Zhang, Ren  \& Sun}{He
  et~al.}{2016}]{he2016deepresiduallearning}
He K.,  Zhang X.,  Ren S.,   Sun J.,  2016, in 2016 IEEE Conference on Computer
  Vision and Pattern Recognition (CVPR). pp 770--778,
  \mn@doi{10.1109/CVPR.2016.90}

\bibitem[\protect\citeauthoryear{{He} et~al.,}{{He} et~al.}{2022}]{he2022}
{He} Q.,  et~al., 2022, \mn@doi [\mnras] {10.1093/mnras/stac191}, \href
  {https://ui.adsabs.harvard.edu/abs/2022MNRAS.511.3046H} {511, 3046}

\bibitem[\protect\citeauthoryear{{Hermans}, {Begy}  \& {Louppe}}{{Hermans}
  et~al.}{2019}]{hermans2019likelihoodfree}
{Hermans} J.,  {Begy} V.,   {Louppe} G.,  2019, arXiv e-prints, \href
  {https://ui.adsabs.harvard.edu/abs/2019arXiv190304057H} {p. arXiv:1903.04057}

\bibitem[\protect\citeauthoryear{{Hermans}, {Banik}, {Weniger}, {Bertone}  \&
  {Louppe}}{{Hermans} et~al.}{2021}]{hermans2021neuralsbi}
{Hermans} J.,  {Banik} N.,  {Weniger} C.,  {Bertone} G.,   {Louppe} G.,  2021,
  \mn@doi [\mnras] {10.1093/mnras/stab2181}, \href
  {https://ui.adsabs.harvard.edu/abs/2021MNRAS.507.1999H} {507, 1999}

\bibitem[\protect\citeauthoryear{Hezaveh, Dalal, Holder, Kisner, Kuhlen  \&
  Perreault~Levasseur}{Hezaveh et~al.}{2016a}]{hezaveh2014measuring}
Hezaveh Y.,  Dalal N.,  Holder G.,  Kisner T.,  Kuhlen M.,
  Perreault~Levasseur L.,  2016a, \mn@doi [JCAP]
  {10.1088/1475-7516/2016/11/048}, 11, 048

\bibitem[\protect\citeauthoryear{{Hezaveh} et~al.,}{{Hezaveh}
  et~al.}{2016b}]{hezaveh2016detection}
{Hezaveh} Y.~D.,  et~al., 2016b, \mn@doi [\apj] {10.3847/0004-637X/823/1/37},
  \href {https://ui.adsabs.harvard.edu/abs/2016ApJ...823...37H} {823, 37}

\bibitem[\protect\citeauthoryear{Hoffer, Hubara  \& Soudry}{Hoffer
  et~al.}{2017}]{hoffer2017train}
Hoffer E.,  Hubara I.,   Soudry D.,  2017, Advances in neural information
  processing systems, 30

\bibitem[\protect\citeauthoryear{{Huang} et~al.,}{{Huang}
  et~al.}{2021}]{huang2021desi}
{Huang} X.,  et~al., 2021, \mn@doi [\apj] {10.3847/1538-4357/abd62b}, \href
  {https://ui.adsabs.harvard.edu/abs/2021ApJ...909...27H} {909, 27}

\bibitem[\protect\citeauthoryear{{Jacobs} et~al.,}{{Jacobs}
  et~al.}{2019}]{jacobs2019des}
{Jacobs} C.,  et~al., 2019, \mn@doi [\apjs] {10.3847/1538-4365/ab26b6}, \href
  {https://ui.adsabs.harvard.edu/abs/2019ApJS..243...17J} {243, 17}

\bibitem[\protect\citeauthoryear{{Kahlhoefer}, {Kaplinghat}, {Slatyer}  \&
  {Wu}}{{Kahlhoefer} et~al.}{2019}]{kahlhoefer2019sidm}
{Kahlhoefer} F.,  {Kaplinghat} M.,  {Slatyer} T.~R.,   {Wu} C.-L.,  2019,
  \mn@doi [\jcap] {10.1088/1475-7516/2019/12/010}, \href
  {https://ui.adsabs.harvard.edu/abs/2019JCAP...12..010K} {2019, 010}

\bibitem[\protect\citeauthoryear{Keeton, Kochanek  \& Seljak}{Keeton
  et~al.}{1997}]{Keeton1996shear}
Keeton C.~R.,  Kochanek C.~S.,   Seljak U.,  1997, \mn@doi [Astrophys. J.]
  {10.1086/304172}, 482, 604

\bibitem[\protect\citeauthoryear{{Kim}, {Peter}  \& {Hargis}}{{Kim}
  et~al.}{2018}]{kim2018missing}
{Kim} S.~Y.,  {Peter} A. H.~G.,   {Hargis} J.~R.,  2018, \mn@doi [\prl]
  {10.1103/PhysRevLett.121.211302}, \href
  {https://ui.adsabs.harvard.edu/abs/2018PhRvL.121u1302K} {121, 211302}

\bibitem[\protect\citeauthoryear{Kingma \& Ba}{Kingma \&
  Ba}{2014}]{kingma2014adam}
Kingma D.~P.,  Ba J.,  2014, arXiv preprint arXiv:1412.6980

\bibitem[\protect\citeauthoryear{{Kochanek} \& {White}}{{Kochanek} \&
  {White}}{2000}]{2000kochanek}
{Kochanek} C.~S.,  {White} M.,  2000, \mn@doi [\apj] {10.1086/317149}, \href
  {https://ui.adsabs.harvard.edu/abs/2000ApJ...543..514K} {543, 514}

\bibitem[\protect\citeauthoryear{{Koekemoer} et~al.,}{{Koekemoer}
  et~al.}{2007}]{koekemoer2007cosmos}
{Koekemoer} A.~M.,  et~al., 2007, \mn@doi [\apjs] {10.1086/520086}, \href
  {https://ui.adsabs.harvard.edu/abs/2007ApJS..172..196K} {172, 196}

\bibitem[\protect\citeauthoryear{{Kormann}, {Schneider}  \&
  {Bartelmann}}{{Kormann} et~al.}{1994}]{kormann1994isothermal}
{Kormann} R.,  {Schneider} P.,   {Bartelmann} M.,  1994, \aap, \href
  {https://ui.adsabs.harvard.edu/abs/1994A&A...284..285K} {284, 285}

\bibitem[\protect\citeauthoryear{{Laureijs} et~al.,}{{Laureijs}
  et~al.}{2011}]{laureijs2011euclid}
{Laureijs} R.,  et~al., 2011, arXiv e-prints, \href
  {https://ui.adsabs.harvard.edu/abs/2011arXiv1110.3193L} {p. arXiv:1110.3193}

\bibitem[\protect\citeauthoryear{Legin, Hezaveh, Levasseur  \& Wandelt}{Legin
  et~al.}{2021}]{Legin:2021zup}
Legin R.,  Hezaveh Y.,  Levasseur L.~P.,   Wandelt B.,  2021, arXiv

\bibitem[\protect\citeauthoryear{{Loshchilov} \& {Hutter}}{{Loshchilov} \&
  {Hutter}}{2017}]{loshchilove2017adamw}
{Loshchilov} I.,  {Hutter} F.,  2017, arXiv e-prints, \href
  {https://ui.adsabs.harvard.edu/abs/2017arXiv171105101L} {p. arXiv:1711.05101}

\bibitem[\protect\citeauthoryear{{Lovell} et~al.,}{{Lovell}
  et~al.}{2012}]{lovell2012wdm}
{Lovell} M.~R.,  et~al., 2012, \mn@doi [\mnras]
  {10.1111/j.1365-2966.2011.20200.x}, \href
  {https://ui.adsabs.harvard.edu/abs/2012MNRAS.420.2318L} {420, 2318}

\bibitem[\protect\citeauthoryear{{Lovell}, {Frenk}, {Eke}, {Jenkins}, {Gao}  \&
  {Theuns}}{{Lovell} et~al.}{2014}]{lovell2014wdm}
{Lovell} M.~R.,  {Frenk} C.~S.,  {Eke} V.~R.,  {Jenkins} A.,  {Gao} L.,
  {Theuns} T.,  2014, \mn@doi [\mnras] {10.1093/mnras/stt2431}, \href
  {https://ui.adsabs.harvard.edu/abs/2014MNRAS.439..300L} {439, 300}

\bibitem[\protect\citeauthoryear{{Lynden-Bell} \& {Wood}}{{Lynden-Bell} \&
  {Wood}}{1968}]{lyndenbell1968}
{Lynden-Bell} D.,  {Wood} R.,  1968, \mn@doi [\mnras]
  {10.1093/mnras/138.4.495}, \href
  {https://ui.adsabs.harvard.edu/abs/1968MNRAS.138..495L} {138, 495}

\bibitem[\protect\citeauthoryear{{MacLeod}, {Jones}, {Agol}  \&
  {Kochanek}}{{MacLeod} et~al.}{2013}]{macleod2013detection}
{MacLeod} C.~L.,  {Jones} R.,  {Agol} E.,   {Kochanek} C.~S.,  2013, \mn@doi
  [\apj] {10.1088/0004-637X/773/1/35}, \href
  {https://ui.adsabs.harvard.edu/abs/2013ApJ...773...35M} {773, 35}

\bibitem[\protect\citeauthoryear{Makinen, Charnock, Alsing  \& Wandelt}{Makinen
  et~al.}{2021}]{Makinen:2021nly}
Makinen T.~L.,  Charnock T.,  Alsing J.,   Wandelt B.~D.,  2021, \mn@doi [JCAP]
  {10.1088/1475-7516/2021/11/049}, 11, 049

\bibitem[\protect\citeauthoryear{{Mandelbaum}, {Hirata}, {Leauthaud}, {Massey}
  \& {Rhodes}}{{Mandelbaum} et~al.}{2012}]{mandelbaum2012precision}
{Mandelbaum} R.,  {Hirata} C.~M.,  {Leauthaud} A.,  {Massey} R.~J.,   {Rhodes}
  J.,  2012, \mn@doi [\mnras] {10.1111/j.1365-2966.2011.20138.x}, \href
  {https://ui.adsabs.harvard.edu/abs/2012MNRAS.420.1518M} {420, 1518}

\bibitem[\protect\citeauthoryear{{Mandelbaum} et~al.,}{{Mandelbaum}
  et~al.}{2014}]{mandelbaum2014great3}
{Mandelbaum} R.,  et~al., 2014, \mn@doi [\apjs] {10.1088/0067-0049/212/1/5},
  \href {https://ui.adsabs.harvard.edu/abs/2014ApJS..212....5M} {212, 5}

\bibitem[\protect\citeauthoryear{{Mao} \& {Schneider}}{{Mao} \&
  {Schneider}}{1998}]{mao1998evidence}
{Mao} S.,  {Schneider} P.,  1998, \mn@doi [\mnras]
  {10.1046/j.1365-8711.1998.01319.x}, \href
  {https://ui.adsabs.harvard.edu/abs/1998MNRAS.295..587M} {295, 587}

\bibitem[\protect\citeauthoryear{{McKean} et~al.,}{{McKean}
  et~al.}{2015}]{mckean2015ska}
{McKean} J.,  et~al., 2015, in Advancing Astrophysics with the Square Kilometre
  Array (AASKA14). p.~84 (\mn@eprint {arXiv} {1502.03362})

\bibitem[\protect\citeauthoryear{{Metcalf} \& {Madau}}{{Metcalf} \&
  {Madau}}{2001}]{metcalf2001compound}
{Metcalf} R.~B.,  {Madau} P.,  2001, \mn@doi [\apj] {10.1086/323695}, \href
  {https://ui.adsabs.harvard.edu/abs/2001ApJ...563....9M} {563, 9}

\bibitem[\protect\citeauthoryear{{Metcalf} et~al.,}{{Metcalf}
  et~al.}{2019}]{metcalf2019strong}
{Metcalf} R.~B.,  et~al., 2019, \mn@doi [\aap] {10.1051/0004-6361/201832797},
  \href {https://ui.adsabs.harvard.edu/abs/2019A&A...625A.119M} {625, A119}

\bibitem[\protect\citeauthoryear{{Minor}, {Kaplinghat}  \& {Li}}{{Minor}
  et~al.}{2017}]{2017ApJ...845..118M}
{Minor} Q.~E.,  {Kaplinghat} M.,   {Li} N.,  2017, \mn@doi [\apj]
  {10.3847/1538-4357/aa7fee}, \href
  {https://ui.adsabs.harvard.edu/abs/2017ApJ...845..118M} {845, 118}

\bibitem[\protect\citeauthoryear{{Minor}, {Kaplinghat}, {Chan}  \&
  {Simon}}{{Minor} et~al.}{2021a}]{minor2021inferring}
{Minor} Q.,  {Kaplinghat} M.,  {Chan} T.~H.,   {Simon} E.,  2021a, \mn@doi
  [\mnras] {10.1093/mnras/stab2209}, \href
  {https://ui.adsabs.harvard.edu/abs/2021MNRAS.507.1202M} {507, 1202}

\bibitem[\protect\citeauthoryear{{Minor}, {Gad-Nasr}, {Kaplinghat}  \&
  {Vegetti}}{{Minor} et~al.}{2021b}]{minor2021unexpected}
{Minor} Q.,  {Gad-Nasr} S.,  {Kaplinghat} M.,   {Vegetti} S.,  2021b, \mn@doi
  [\mnras] {10.1093/mnras/stab2247}, \href
  {https://ui.adsabs.harvard.edu/abs/2021MNRAS.507.1662M} {507, 1662}

\bibitem[\protect\citeauthoryear{{Mishra-Sharma}}{{Mishra-Sharma}}{2022}]{mishrasharma2022inferring}
{Mishra-Sharma} S.,  2022, \mn@doi [Machine Learning: Science and Technology]
  {10.1088/2632-2153/ac494a}, \href
  {https://ui.adsabs.harvard.edu/abs/2022MLS&T...3aLT03M} {3, 01LT03}

\bibitem[\protect\citeauthoryear{{Mishra-Sharma}, {Van Tilburg}  \&
  {Weiner}}{{Mishra-Sharma} et~al.}{2020}]{mishrasharma2020powerhalometry}
{Mishra-Sharma} S.,  {Van Tilburg} K.,   {Weiner} N.,  2020, \mn@doi [\prd]
  {10.1103/PhysRevD.102.023026}, \href
  {https://ui.adsabs.harvard.edu/abs/2020PhRvD.102b3026M} {102, 023026}

\bibitem[\protect\citeauthoryear{{Mohamed} \& {Lakshminarayanan}}{{Mohamed} \&
  {Lakshminarayanan}}{2016}]{mohamed2016learning}
{Mohamed} S.,  {Lakshminarayanan} B.,  2016, arXiv e-prints, \href
  {https://ui.adsabs.harvard.edu/abs/2016arXiv161003483M} {p. arXiv:1610.03483}

\bibitem[\protect\citeauthoryear{{Mondino}, {Taki}, {Van Tilburg}  \&
  {Weiner}}{{Mondino} et~al.}{2020}]{mondino2020first}
{Mondino} C.,  {Taki} A.-M.,  {Van Tilburg} K.,   {Weiner} N.,  2020, \mn@doi
  [\prl] {10.1103/PhysRevLett.125.111101}, \href
  {https://ui.adsabs.harvard.edu/abs/2020PhRvL.125k1101M} {125, 111101}

\bibitem[\protect\citeauthoryear{{Moustakas} \& {Metcalf}}{{Moustakas} \&
  {Metcalf}}{2003}]{moustakas2003detecting}
{Moustakas} L.~A.,  {Metcalf} R.~B.,  2003, \mn@doi [\mnras]
  {10.1046/j.1365-8711.2003.06055.x}, \href
  {https://ui.adsabs.harvard.edu/abs/2003MNRAS.339..607M} {339, 607}

\bibitem[\protect\citeauthoryear{{Navarro}, {Frenk}  \& {White}}{{Navarro}
  et~al.}{1997}]{navarro1997nfw}
{Navarro} J.~F.,  {Frenk} C.~S.,   {White} S. D.~M.,  1997, \mn@doi [\apj]
  {10.1086/304888}, \href
  {https://ui.adsabs.harvard.edu/abs/1997ApJ...490..493N} {490, 493}

\bibitem[\protect\citeauthoryear{{Nierenberg}, {Treu}, {Wright}, {Fassnacht}
  \& {Auger}}{{Nierenberg} et~al.}{2014}]{nierenberg2014detection}
{Nierenberg} A.~M.,  {Treu} T.,  {Wright} S.~A.,  {Fassnacht} C.~D.,   {Auger}
  M.~W.,  2014, \mn@doi [\mnras] {10.1093/mnras/stu862}, \href
  {https://ui.adsabs.harvard.edu/abs/2014MNRAS.442.2434N} {442, 2434}

\bibitem[\protect\citeauthoryear{{Nierenberg} et~al.,}{{Nierenberg}
  et~al.}{2017}]{nierenberg2017probing}
{Nierenberg} A.~M.,  et~al., 2017, \mn@doi [\mnras] {10.1093/mnras/stx1400},
  \href {https://ui.adsabs.harvard.edu/abs/2017MNRAS.471.2224N} {471, 2224}

\bibitem[\protect\citeauthoryear{Ostdiek, Diaz~Rivero  \& Dvorkin}{Ostdiek
  et~al.}{2022a}]{Ostdiek:2020cqz}
Ostdiek B.,  Diaz~Rivero A.,   Dvorkin C.,  2022a, \mn@doi [Astron. Astrophys.]
  {10.1051/0004-6361/202142030}, 657, L14

\bibitem[\protect\citeauthoryear{Ostdiek, Diaz~Rivero  \& Dvorkin}{Ostdiek
  et~al.}{2022b}]{Ostdiek:2020mvo}
Ostdiek B.,  Diaz~Rivero A.,   Dvorkin C.,  2022b, \mn@doi [Astrophys. J.]
  {10.3847/1538-4357/ac2d8d}, 927, 83

\bibitem[\protect\citeauthoryear{Paszke et~al.,}{Paszke
  et~al.}{2019}]{NEURIPS2019_9015}
Paszke A.,  et~al., 2019, in Wallach H.,  Larochelle H.,  Beygelzimer A.,
  d\textquotesingle Alch\'{e}-Buc F.,  Fox E.,   Garnett R.,  eds, , Advances
  in Neural Information Processing Systems 32.
Curran Associates, Inc., pp 8024--8035, \url
  {http://papers.neurips.cc/paper/9015-pytorch-an-imperative-style-high-performance-deep-learning-library.pdf}

\bibitem[\protect\citeauthoryear{Perreault~Levasseur, Hezaveh  \&
  Wechsler}{Perreault~Levasseur
  et~al.}{2017}]{PerreaultLevasseur2017uncertainties}
Perreault~Levasseur L.,  Hezaveh Y.~D.,   Wechsler R.~H.,  2017, \mn@doi
  [Astrophys. J. Lett.] {10.3847/2041-8213/aa9704}, 850, L7

\bibitem[\protect\citeauthoryear{{Planck Collaboration} et~al.,}{{Planck
  Collaboration} et~al.}{2020}]{planck2020planck}
{Planck Collaboration} et~al., 2020, \mn@doi [\aap]
  {10.1051/0004-6361/201833880}, \href
  {https://ui.adsabs.harvard.edu/abs/2020A&A...641A...1P} {641, A1}

\bibitem[\protect\citeauthoryear{{Read}, {Iorio}, {Agertz}  \&
  {Fraternali}}{{Read} et~al.}{2017}]{read2017stellar}
{Read} J.~I.,  {Iorio} G.,  {Agertz} O.,   {Fraternali} F.,  2017, \mn@doi
  [\mnras] {10.1093/mnras/stx147}, \href
  {https://ui.adsabs.harvard.edu/abs/2017MNRAS.467.2019R} {467, 2019}

\bibitem[\protect\citeauthoryear{{Ritondale}, {Vegetti}, {Despali}, {Auger},
  {Koopmans}  \& {McKean}}{{Ritondale} et~al.}{2019}]{ritondale2019lowmass}
{Ritondale} E.,  {Vegetti} S.,  {Despali} G.,  {Auger} M.~W.,  {Koopmans}
  L.~V.~E.,   {McKean} J.~P.,  2019, \mn@doi [\mnras] {10.1093/mnras/stz464},
  \href {https://ui.adsabs.harvard.edu/abs/2019MNRAS.485.2179R} {485, 2179}

\bibitem[\protect\citeauthoryear{{Rocha}, {Peter}, {Bullock}, {Kaplinghat},
  {Garrison-Kimmel}, {O{\~n}orbe}  \& {Moustakas}}{{Rocha}
  et~al.}{2013}]{rocha2013sidm}
{Rocha} M.,  {Peter} A. H.~G.,  {Bullock} J.~S.,  {Kaplinghat} M.,
  {Garrison-Kimmel} S.,  {O{\~n}orbe} J.,   {Moustakas} L.~A.,  2013, \mn@doi
  [\mnras] {10.1093/mnras/sts514}, \href
  {https://ui.adsabs.harvard.edu/abs/2013MNRAS.430...81R} {430, 81}

\bibitem[\protect\citeauthoryear{{Scoville} et~al.,}{{Scoville}
  et~al.}{2007}]{scoville2007cosmos}
{Scoville} N.,  et~al., 2007, \mn@doi [\apjs] {10.1086/516585}, \href
  {https://ui.adsabs.harvard.edu/abs/2007ApJS..172....1S} {172, 1}

\bibitem[\protect\citeauthoryear{{Spergel} \& {Steinhardt}}{{Spergel} \&
  {Steinhardt}}{2000}]{spergel2000sidm}
{Spergel} D.~N.,  {Steinhardt} P.~J.,  2000, \mn@doi [\prl]
  {10.1103/PhysRevLett.84.3760}, \href
  {https://ui.adsabs.harvard.edu/abs/2000PhRvL..84.3760S} {84, 3760}

\bibitem[\protect\citeauthoryear{{Storfer} et~al.,}{{Storfer}
  et~al.}{2022}]{storfer2022desi}
{Storfer} C.,  et~al., 2022, arXiv e-prints, \href
  {https://ui.adsabs.harvard.edu/abs/2022arXiv220602764S} {p. arXiv:2206.02764}

\bibitem[\protect\citeauthoryear{Taylor, Kitching, Alsing, Wandelt, Feeney  \&
  McEwen}{Taylor et~al.}{2019}]{delfi2}
Taylor P.~L.,  Kitching T.~D.,  Alsing J.,  Wandelt B.~D.,  Feeney S.~M.,
  McEwen J.~D.,  2019, \mn@doi [Physical Review D]
  {10.1103/physrevd.100.023519}, 100

\bibitem[\protect\citeauthoryear{{Van Tilburg}, {Taki}  \& {Weiner}}{{Van
  Tilburg} et~al.}{2018}]{vantilburg2018halometry}
{Van Tilburg} K.,  {Taki} A.-M.,   {Weiner} N.,  2018, \mn@doi [\jcap]
  {10.1088/1475-7516/2018/07/041}, \href
  {https://ui.adsabs.harvard.edu/abs/2018JCAP...07..041V} {2018, 041}

\bibitem[\protect\citeauthoryear{{Vegetti}, {Koopmans}, {Bolton}, {Treu}  \&
  {Gavazzi}}{{Vegetti} et~al.}{2010}]{vegetti2010detection}
{Vegetti} S.,  {Koopmans} L.~V.~E.,  {Bolton} A.,  {Treu} T.,   {Gavazzi} R.,
  2010, \mn@doi [\mnras] {10.1111/j.1365-2966.2010.16865.x}, \href
  {https://ui.adsabs.harvard.edu/abs/2010MNRAS.408.1969V} {408, 1969}

\bibitem[\protect\citeauthoryear{{Vegetti}, {Lagattuta}, {McKean}, {Auger},
  {Fassnacht}  \& {Koopmans}}{{Vegetti} et~al.}{2012}]{vegetti2012detection}
{Vegetti} S.,  {Lagattuta} D.~J.,  {McKean} J.~P.,  {Auger} M.~W.,  {Fassnacht}
  C.~D.,   {Koopmans} L.~V.~E.,  2012, \mn@doi [\nat] {10.1038/nature10669},
  \href {https://ui.adsabs.harvard.edu/abs/2012Natur.481..341V} {481, 341}

\bibitem[\protect\citeauthoryear{{Vegetti}, {Koopmans}, {Auger}, {Treu}  \&
  {Bolton}}{{Vegetti} et~al.}{2014}]{vegetti2014inferencecdmsubmf}
{Vegetti} S.,  {Koopmans} L.~V.~E.,  {Auger} M.~W.,  {Treu} T.,   {Bolton}
  A.~S.,  2014, \mn@doi [\mnras] {10.1093/mnras/stu943}, \href
  {https://ui.adsabs.harvard.edu/abs/2014MNRAS.442.2017V} {442, 2017}

\bibitem[\protect\citeauthoryear{{Vogelsberger}, {Zavala}  \&
  {Loeb}}{{Vogelsberger} et~al.}{2012}]{vogelsberger2012sidm}
{Vogelsberger} M.,  {Zavala} J.,   {Loeb} A.,  2012, \mn@doi [\mnras]
  {10.1111/j.1365-2966.2012.21182.x}, \href
  {https://ui.adsabs.harvard.edu/abs/2012MNRAS.423.3740V} {423, 3740}

\bibitem[\protect\citeauthoryear{Wagner-Carena, Park, Birrer, Marshall, Roodman
   \& Wechsler}{Wagner-Carena et~al.}{2021}]{Wagner-Carena2020hierarchical}
Wagner-Carena S.,  Park J.~W.,  Birrer S.,  Marshall P.~J.,  Roodman A.,
  Wechsler R.~H.,  2021, \mn@doi [Astrophys. J.] {10.3847/1538-4357/abdf59},
  909, 187

\bibitem[\protect\citeauthoryear{{Wagner-Carena}, {Aalbers}, {Birrer},
  {Nadler}, {Darragh-Ford}, {Marshall}  \& {Wechsler}}{{Wagner-Carena}
  et~al.}{2022}]{carenawagner2022images}
{Wagner-Carena} S.,  {Aalbers} J.,  {Birrer} S.,  {Nadler} E.~O.,
  {Darragh-Ford} E.,  {Marshall} P.~J.,   {Wechsler} R.~H.,  2022, arXiv
  e-prints, \href {https://ui.adsabs.harvard.edu/abs/2022arXiv220300690W} {p.
  arXiv:2203.00690}

\bibitem[\protect\citeauthoryear{Wilks}{Wilks}{1938}]{Wilks:1938dza}
Wilks S.~S.,  1938, \mn@doi [Annals Math. Statist.] {10.1214/aoms/1177732360},
  9, 60

\bibitem[\protect\citeauthoryear{{{\c{S}}eng{\"u}l} \&
  {Dvorkin}}{{{\c{S}}eng{\"u}l} \& {Dvorkin}}{2022}]{sengul2022probing}
{{\c{S}}eng{\"u}l} A.~{\c{C}}.,  {Dvorkin} C.,  2022, \mn@doi [\mnras]
  {10.1093/mnras/stac2256}, \href
  {https://ui.adsabs.harvard.edu/abs/2022MNRAS.516..336S} {516, 336}

\bibitem[\protect\citeauthoryear{\c{S}eng\"ul, Tsang, Diaz~Rivero, Dvorkin, Zhu
   \& Seljak}{\c{S}eng\"ul et~al.}{2020}]{CaganSengul:2020nat}
\c{S}eng\"ul A.~C.,  Tsang A.,  Diaz~Rivero A.,  Dvorkin C.,  Zhu H.-M.,
  Seljak U.,  2020, \mn@doi [Phys. Rev. D] {10.1103/PhysRevD.102.063502}, 102,
  063502

\makeatother
\end{thebibliography}

\label{lastpage}

\end{document}